\documentclass[submission, Phys]{SciPost}
\hypersetup{
    colorlinks,
    linkcolor={red!50!black},
    citecolor={blue!50!black},
    urlcolor={blue!80!black}
}

\usepackage[bitstream-charter]{mathdesign}
\DeclareSymbolFont{usualmathcal}{OMS}{cmsy}{m}{n}
\DeclareSymbolFontAlphabet{\mathcal}{usualmathcal}

\usepackage{enumitem}
\setlist[enumerate]{
  label=(\roman*), 
  itemsep=0\baselineskip,
  topsep=0.125\baselineskip,
  labelwidth=1.5em,       % narrower label box
  labelsep=0.5em,         % smaller space between label and text
  leftmargin=!            % compute left margin from labelwidth+labelsep
}

\usepackage[normalem]{ulem}
\usepackage{lmodern}
\usepackage{ytableau}
\usepackage{pifont}
\usepackage{color}
\usepackage{bm}
\usepackage{wasysym}

\usepackage{ dsfont }
\usepackage{graphicx}
\usepackage{algorithm}
\usepackage{algorithmic}
\usepackage{tikz}

\begin{document}

\begin{center}{\Large \textbf{
Extracting average properties of disordered spin chains with translationally invariant tensor networks\\
}}\end{center}

% TODO: write the author list here. Use first name (+ other initials) + surname format.
% Separate subsequent authors by a comma, omit comma and use "and" for the last author.
% Mark the corresponding author with a superscript star.
\begin{center}
Kevin Vervoort\textsuperscript{1},
Wei Tang\textsuperscript{1} and
Nick Bultinck\textsuperscript{1}
\end{center}

\begin{center}
{\bf 1} Department of Physics and Astronomy, Ghent University, Krijgslaan 281, 9000 Gent, Belgium
\\
${}^\star$ {\small \sf Kevin.vervoort@Ugent.be}
\end{center}

\begin{center}
\today
\end{center}

% For convenience during refereeing (optional),
% you can turn on line numbers by uncommenting the next line:
%\linenumbers

\section*{Abstract}
{\bf
% TODO: write your abstract here.
We develop a tensor network-based method for calculating disorder-averaged finite-temperature expectation values in random spin chains without having to explicitly sample over disorder configurations. The algorithm exploits statistical translation invariance and works directly in the thermodynamic limit. We benchmark our method on the infinite-randomness critical point of the random transverse field Ising model.
}

% \vspace{10pt}
% \noindent\rule{\textwidth}{1pt}
% \tableofcontents\thispagestyle{fancy}
% \noindent\rule{\textwidth}{1pt}
% \vspace{10pt}

\section{Introduction}
\label{sec:intro}
% TODO: write your article here.
Since the invention of the Density Matrix Renormalization Group (DMRG) algorithm in 1992~\cite{White1992}, a broad array of powerful tensor network-based methods has been developed for studying quantum many-body systems (see e.g.~\cite{Schollwock2005,Verstraete2008,Vanderstraeten2019,Banuls2023,xiang-density-book-2023} and references therein). The community has reached a point at which low-temperature equilibrium properties of most local spin and fermion Hamiltonians on a one-dimensional lattice can be studied with rather limited computational resources, including many gapless systems~\cite{Schollwock2005,xiang-density-book-2023}. One notable exception, however, are systems with quenched randomness. The challenge in simulating these systems is two-fold. First, there is no translation symmetry such that the left-right sweeping procedure of the original DMRG algorithm appears to be the most appropriate way to optimize the Matrix Product State (MPS). But randomness can cause entanglement to be inhomogeneously distributed on different length scales (such as for example in random singlet states~\cite{Ma1979,Dasgupta1980,Bhatt1982}), which significantly increases the number of sweeps needed for convergence. Secondly, calculating disorder-averaged properties requires a large number of simulations with different disorder samples. Some interesting recent progress on the first challenge has been made via a numerical implementation of the rigorous renormalization group~\cite{Roberts2017,Roberts2021}, but the second challenge in dealing with a large number of disorder samples remains.

In this work we address this issue for finite temperature properties of random spin chains. We consider systems with statistical translation invariance, meaning that the distribution of the random variables is the same on every site, and represent the density matrix from which disorder-averaged expectation values are obtained as a Matrix Product Operator (MPO)~\cite{Haegeman2017}. MPOs are a straightforward generalization of MPS to operators, and by exploiting (statistical) translation invariance they can also be used directly in the thermodynamic limit while still permitting an efficient evaluation of expectation values. We present an algorithm to explicitly construct the desired MPO starting from the infinite-temperature state, and illustrate it on the random transverse field Ising model near criticality. The low temperature properties of this model have been studied numerically~\cite{young1996numerical,young1997finite} and will provide a nontrivial benchmark for our algorithm. Down to low temperatures we can reproduce the average properties of this paradigmatic model with relatively small bond dimension (order $\sim 100$).

\section{Algorithm}
\label{sec:algorithm}
We consider disordered Hamiltonians of the form $\mathcal{H}[\{R_n\}]=\sum_n \tilde{h}_n[R_n]$, where the integer $n$ labels the lattice sites and the local terms $\tilde{h}_n[R_n]$ act non-trivially on a finite number of consecutive spins starting at $n$. The local Hamiltonian terms depend on the quenched disorder variables $R_n$. In what follows we consider systems with discrete disorder where the disorder parameters $R_n$ can take on $N_D$ different values, with probability $P(R_n)$. We take $P(R_n)$ to be the same at every site, leading to statistical translation invariance. The main idea behind our algorithm is to approximate the mixture of Gibbs states with different quenched disorder configurations by a single MPO, which is translationally invariant due to the statistical translation invariance. Concretely, we devise an algorithm to approximate following density matrix:
\begin{equation}
\rho = \sum_{\{R_n\}}\left[\prod_n P(R_n)\right] \rho_G[\{R_n\}]\,,\label{defrho}
\end{equation}
where the sum is over disorder configurations. The density matrices $\rho_G$ are Gibbs states for the different disorder realizations:
\begin{equation}
\rho_G[\{R_n\}] = e^{-H[\{R_n\}]/T}/Z[\{R_n\}]\,,
\end{equation}
with $T$ the temperature and $Z[\{R_n\}] = \text{tr}\left(e^{-H[\{R_n\}]/T} \right)$ the partition functions. It follows from the definition of $\rho$ that $\text{tr}\left(\rho O\right)$ is the disorder-averaged thermal expectation value of $O$. Before going into the details of the algorithm, let us note that (in contrast to clean spin chains~\cite{Hastings2006,Molnar2015}) there is no a priori theoretical reason to assume that $\rho$ has an efficient MPO approximation. One of the main results of this work is therefore to provide numerical evidence that such an efficient approximation does indeed exist, and can capture the non-trivial physics of disordered spin chains.

As in Refs.~\cite{Paredes2005,hubig-time-dependent-2019}, which studied real-time evolution for disorder systems, we introduce ancilla qudits $|R_n\rangle$ of dimension $N_D$ on every site, and define a translationally invariant Hamiltonian $H=\sum_n h_n$, where the local terms $h_n$ are
\begin{equation}
h_n = \sum_{R_n=1}^{N_D}\tilde{h}_n[R_n]\otimes| R_n\rangle \langle R_n|
\end{equation}
These terms represent a controlled action of the original terms $\tilde{h}_n[R_n]$ in the disordered Hamiltonian, where the ancilla qudits play the role of the control qudits. As with controlled quantum gates, the Hamiltonian terms $h_n$ are diagonal in the control (or disorder) qudits, and simply measure their value. Depending on this value a particular local Hamiltonian term of the disordered Hamiltonian acts on the physical spins.

It is important to note that $\rho$ as defined in Eq.~\eqref{defrho} is \emph{not} simply the Gibbs state of $H$ with traced out disorder qudits. This is because first constructing the Gibbs state of $H$ and then tracing out the disorder qudits fails to include the all-important normalization factors $1/Z[\{R_n\}]$. The algorithm we develop in this work is therefore a modified version of Time-Evolving Block Decimation (TEBD)~\cite{Vidal2003}, which constructs the Gibbs state of $H$ via imaginary-time evolution while at the same time also ensuring the correct normalization.

During the imaginary-time evolution, we work with following operator:
\begin{equation}
\tilde{\rho}(\tau) = N(\tau)e^{-\tau H}\,,\label{eq:rho_tilde}
\end{equation}
where $N(\tau) = N_R(\tau)\otimes \mathds{1}_{\sigma}$ is a diagonal operator acting as the identity on the physical spins, and whose diagonal elements are given by $1/Z_\tau[\{R_n\}]$, i.e. the inverse partition functions at imaginary time $\tau$ for the different disorder configurations. The operator $N(\tau)$ thus ensures that tracing out the physical spins $\sigma_n$ in $\tilde{\rho}(\tau)$ produces an identity matrix on the disorder qudits: $\text{tr}_{\{\sigma\}}\tilde{\rho}(\tau) = \mathds{1}_R$.
\begin{figure}
\begin{center}
        \includegraphics[scale=0.75]{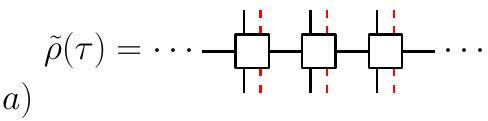} \\
        \vspace{0.25 cm}
        \includegraphics[scale=0.75]{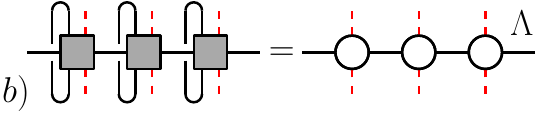} \\
        \vspace{0.25 cm}
        \includegraphics[scale=0.75]{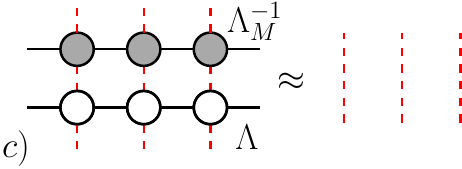} 
        \end{center}
        \caption{(a) MPO representation of $\tilde{\rho}(\tau)$. Horizontal lines represent the virtual bonds. Vertical black full lines represent the physical spin indices. Red dashed lines act on the disorder qudits. (b) The MPO $\Lambda$ is obtained by tracing out the physical spin indices in $N(\tau)e^{-(\tau+\Delta \tau)H}$. (c) An ansatz MPO $\Lambda^{-1}_M$ (filled circles) is used to approximately invert $\Lambda$.} \label{fig:MPOs}
\end{figure}

At every step $\tilde{\rho}(\tau)$ will be represented as an infinite MPO [see Fig.~\ref{fig:MPOs} (a)]:
\begin{align}
\tilde{\rho}&(\tau) = \sum_{\{\sigma\},\{\sigma'\}}\sum_{\{R\}} \text{tr}\left(\cdots A^{\sigma_{n},\sigma'_n}_{R_n}A^{\sigma_{n+1},\sigma'_{n+1}}_{R_{n+1}}\cdots \right) \\
& \cdots |\sigma_{n},R_n\rangle\langle \sigma'_{n},R_n|\otimes |\sigma_{n+1},R_{n+1}\rangle\langle \sigma'_{n+1},R_{n+1}| \cdots \,,\nonumber
\end{align}
where the $D\times D$ matrices $A^{\sigma_n,\sigma'_n}_{R_n}$ are the same on every site.
 Here, $\sigma_n$ and $\sigma'_n$ are the physical spin indices (vertical black lines in Fig.~\ref{fig:MPOs} (a)), and $R_n$ is the disorder qudit index (red dashed lines in Fig.~\ref{fig:MPOs} (a)). 

We start at $\tau=0$, for which $\tilde{\rho}(0) \propto \mathds{1}$, and construct an MPO approximation for $\tilde{\rho}(\beta)$ by iteratively going through following 3 steps: (1) imaginary time-evolve with $H$ from $\tau$ to $\tau+\Delta \tau$, (2) adjust the normalization operator $N(\tau) \rightarrow N(\tau+\Delta \tau)$, and (3) truncate the virtual bond dimension of the MPO. 
After step (3), we redefine $\tau + \Delta \tau$ as $\tau$, and go back to step (1). This cycle is iterated until a particular imaginary time $\beta = T^{-1}$ is reached, after which we directly obtain an approximation for the density matrix in Eq.~\eqref{defrho}: \begin{equation}
\rho \approx \text{tr}_{\{R\}}\left(\left[\prod_n P(R_n) \right] \tilde{\rho}(\beta)\right)    
\end{equation}

Let us now explain these 3 steps in more detail. The first step consists of a conventional TEBD update based on a Suzuki-Trotter approximation of $e^{-\Delta \tau H}$. As this step is standard (i.e. it is identical to a TEBD step for clean spin chains), we will not explain it in detail here. Let us only mention that the bond dimension of the MPO increases from $D$ to $DD_{T}$ during the TEBD update, with $D_{T}$ the bond dimension of the Trotter gates. 

 After the first step, the MPO corresponds to $N(\tau)e^{-(\tau+\Delta \tau)H}$, which deviates from the desired normalized form in Eq.~\eqref{eq:rho_tilde}. 
To update the normalization operator { from $N(\tau)$ to $N(\tau+\Delta\tau)$}, in the second step, 
 
we trace out the physical spins to obtain 
\begin{equation}
\text{tr}_{\{\sigma\}}\left(N(\tau)e^{-(\tau+\Delta \tau)H} \right) = N_R(\tau)N_R^{-1}(\tau+\Delta \tau) =: \Lambda\,,
\end{equation}
which is a diagonal MPO acting on the disorder qudits. The MPO $\Lambda$ is represented graphically in Fig.~\ref{fig:MPOs} (b). In practice, we find that the bond dimension of $\Lambda$, which is also $DD_T$, is highly redundant and can be truncated down to a much smaller value without any significant loss of precision. We now want to find an MPO representation of $\Lambda^{-1} = N_R^{-1}(\tau)N_R(\tau+\Delta \tau)$. 
We do this variationally by fixing a bond dimension $\chi$ of an ansatz MPO $\Lambda^{-1}_M$, which we take to be diagonal in the physical indices, and maximizing the fidelity between $\Lambda \Lambda^{-1}_M$ and the identity operator $\mathds{1}_R$ [see Fig.~\ref{fig:MPOs} (c)]. 
This can be done directly in the thermodynamic limit by a straightforward generalization of the VOMPS algorithm (variational optimization of matrix product state)~\cite{haegeman-post-2013,Vanderstraeten2019,Vanhecke2021}, which maximizes the fidelity per site. If the desired tolerance for minimizing the cost function is not achieved, we increase $\chi$. More details on the modified VOMPS algorithm can be found in the appendix. At the end of step $(2)$ we obtain 
\begin{equation}
\tilde{\rho}(\tau + \Delta \tau) = N(\tau+\Delta \tau)e^{-(\tau+\Delta \tau)H} \approx (\Lambda^{-1}_M\otimes \mathds{1}_\sigma) N(\tau) e^{-(\tau+\Delta \tau)H}\, ,
\end{equation}
which is an MPO of bond dimension $DD_T\chi$. 

In the third and final step, we truncate this MPO back to an MPO of bond dimension $D$.
 In this step, we perform a standard TEBD truncation of $\tilde{\rho}(\tau+\Delta \tau)$, where the MPO is truncated according to the Schmidt values of a bipartition of the system into a left- and right-infinite half.

Compared to the imaginary-time evolution of clean spin chains, the most important new step in our algorithm is to adjust the normalization $N(\tau)$ and the corresponding VOMPS step to find the MPO approximation of $\Lambda^{-1}$. 
Such normalization steps were not required in previous works focusing on real-time dynamics of disordered spin chains using ancilla qudits, as the unitary real-time evolution does not change the normalization of the density matrix~\cite{Paredes2005,hubig-time-dependent-2019}. 
However, they are crucial for our algorithm. Specifically, we expect that  it is crucial to first restore the normalization before truncating the MPO. Otherwise, the truncation would not treat the different disorder sectors equally and could discard important contributions to $\rho$. 
 Let us also note that for sufficiently small $\Delta \tau$, $\Lambda = N_R^{-1}(\tau)N_R(\tau+\Delta \tau)$ is close to the identity, which has a trivial (MPO) inverse. Taking $\Delta \tau$ small enough, it should therefore always be possible to approximately invert $\Lambda$ with a low bond-dimension MPO.
%We also expect that after every imaginary-time step it is crucial to first restore the normalization before truncating the MPO. Otherwise, the truncation would not treat the different disorder sectors equally and could discard important contributions to $\rho$. 

The dominant numerical cost of the algorithm comes from truncating the MPO density matrix.
Compared to the truncation procedure for clean systems, this cost is higher in our algorithm due to the additional ancilla qubits and the additional normalization step.
The ancilla qubits increase the dimension of the open indices of the MPO, thereby increasing the truncation cost by a factor of $N_D$ (note that the MPO is diagonal in the ancilla degrees of freedom, resulting in a linear-in-$N_D$ memory growth.
The normalization step increases the MPO virtual bond dimension by a factor of $\chi$, which increases the truncation cost by a factor of $\chi^3$. This increase in bond dimension makes that the truncation step, which involves finding the leading eigenvector of the transfer matrix, becomes the most costly step in both runtime and memory. 
In practice, we find that the approximate MPO inversion can be performed with a relatively small $\chi$ ($\lesssim 8$), which does not appear to increase significantly with the number of disorder values used in the simulations (based on the simulations for the random transverse-field Ising model presented in the next section).
One can also take smaller imaginary-time steps, which, as discussed above, should reduce the $\chi$ required for accurate inversion, at the expense of a linear increase in the runtime needed to reach a desired temperature.

\section{Application to Random Transverse Field Ising Model}
\label{sec:RTFIM}

\begin{figure}
    \includegraphics[scale=0.3]{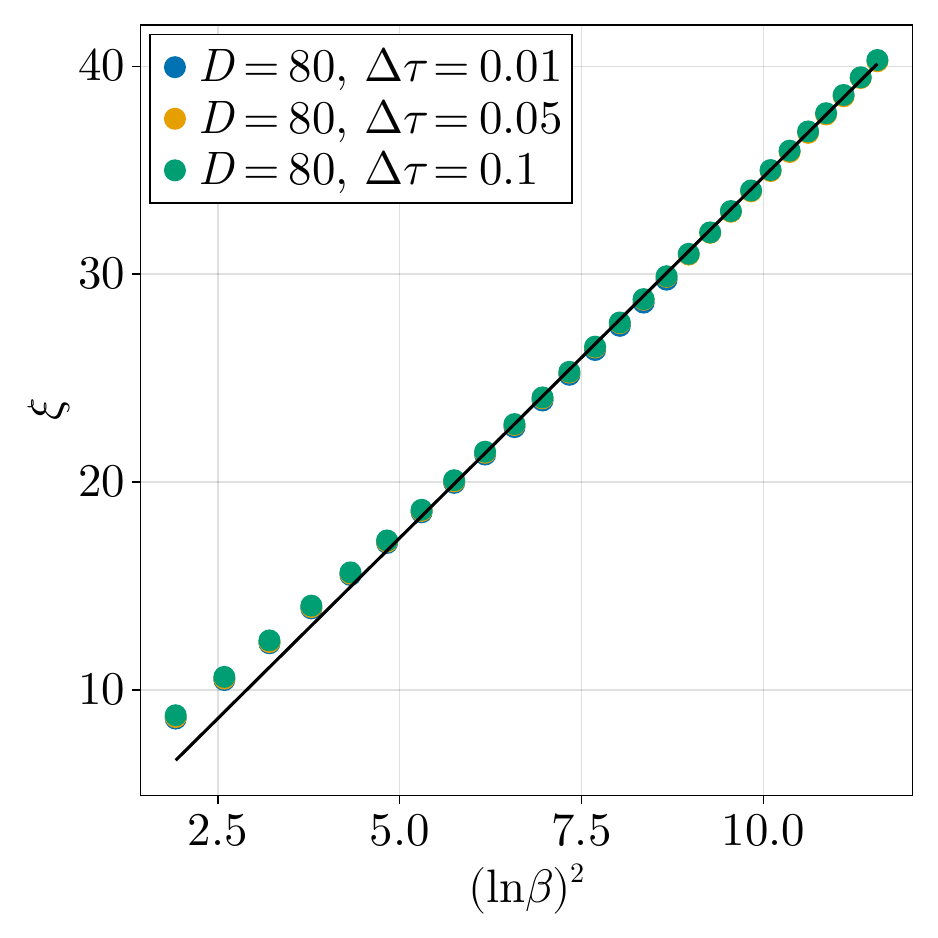} 
    \includegraphics[scale=0.3]{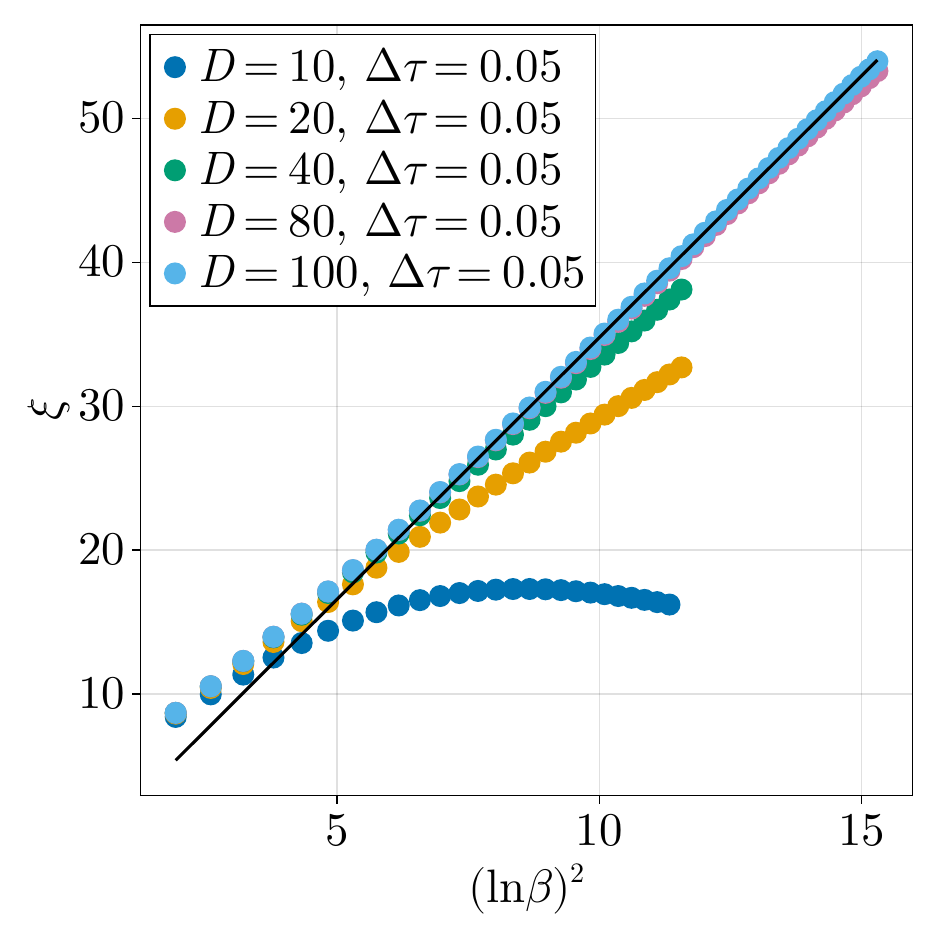} 
    \includegraphics[scale=0.3]{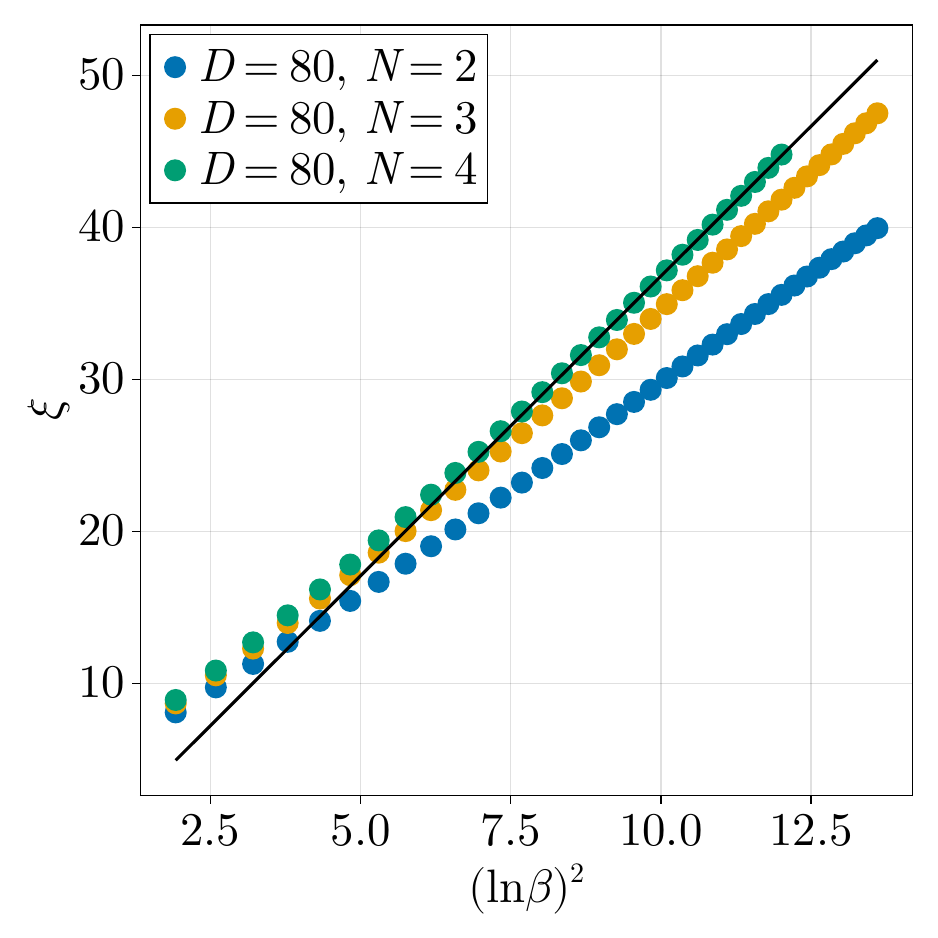} 
    \caption{Left: Correlation length of the average correlation function as a function of $(\ln\beta)^2$ for different step sizes $\Delta \tau$. The black line corresponds to a linear fit of the data for $\Delta\tau = 0.1$ and $\beta \in [10,30]$. Middle: Correlation length as a function of $(\ln\beta)^2$ for different bond dimensions. The black line represents a linear fit of the data for $D = 100$ and $\beta \in [30,50]$. Right: Correlation length as a function $(\ln\beta)^2$ for different numbers of disorder values $N_D = N^2$, obtained with $\Delta\tau = 0.05$. The black line represents a linear fit of the data with $N = 4$ and $\beta \in [20,30]$. In all figures the correlation length was extracted from the transfer matrix eigenvalues.} \label{fig:corr_length}
\end{figure}

We now illustrate our algorithm by applying it to the solvable~\cite{WuMcCoy,Fisher1992,Fisher1995, igloi2005strong} random transverse field Ising model (RTFIM). The Hamiltonian is given by
\begin{equation}
    \mathcal{H}_{\mathrm{I}} = -\sum_n J_n \sigma^z_n\sigma^z_{n+1} + \sum_n h_n\sigma^x_n\,,
\end{equation}
where $J_n$ and $h_n$ are uncorrelated random variables. For our simulations we take both $J_n$ and $h_n$ to be uniformly distributed with values $[0.7, 1.0, 1.3]$,  resulting in a total of $N_D=9$ disorder values. This distribution has
\begin{equation}
\delta = \frac{\langle\ln h_n\rangle-\langle\ln J_n\rangle}{\text{Var}(\ln h_n)+\text{Var}(\ln J_n)} = 0\,,
\end{equation}
which means it is at the $\mathbb{Z}_2$-breaking critical point ~\cite{Fisher1992,Fisher1995}, and is sufficiently broad to erase the clean Ising behavior already at short length scales~\cite{Laflorencie2004}. We consider this model to be a highly non-trivial test for our algorithm, as the RTFIM has an \emph{infinite-randomness} critical point where disorder-averaged quantities receive important contributions from rare regions~\cite{Fisher1992}. For the numerical simulations we set the tolerance for the inverse of $\Lambda$ at $10^{-6}$. The bond dimension of $\Lambda$ was truncated to $4$, hence we required only a bond dimension of $\chi =2$ or $\chi=4$ for $\Lambda^{-1}_M$. 
\begin{figure}
    \includegraphics[scale=0.33]{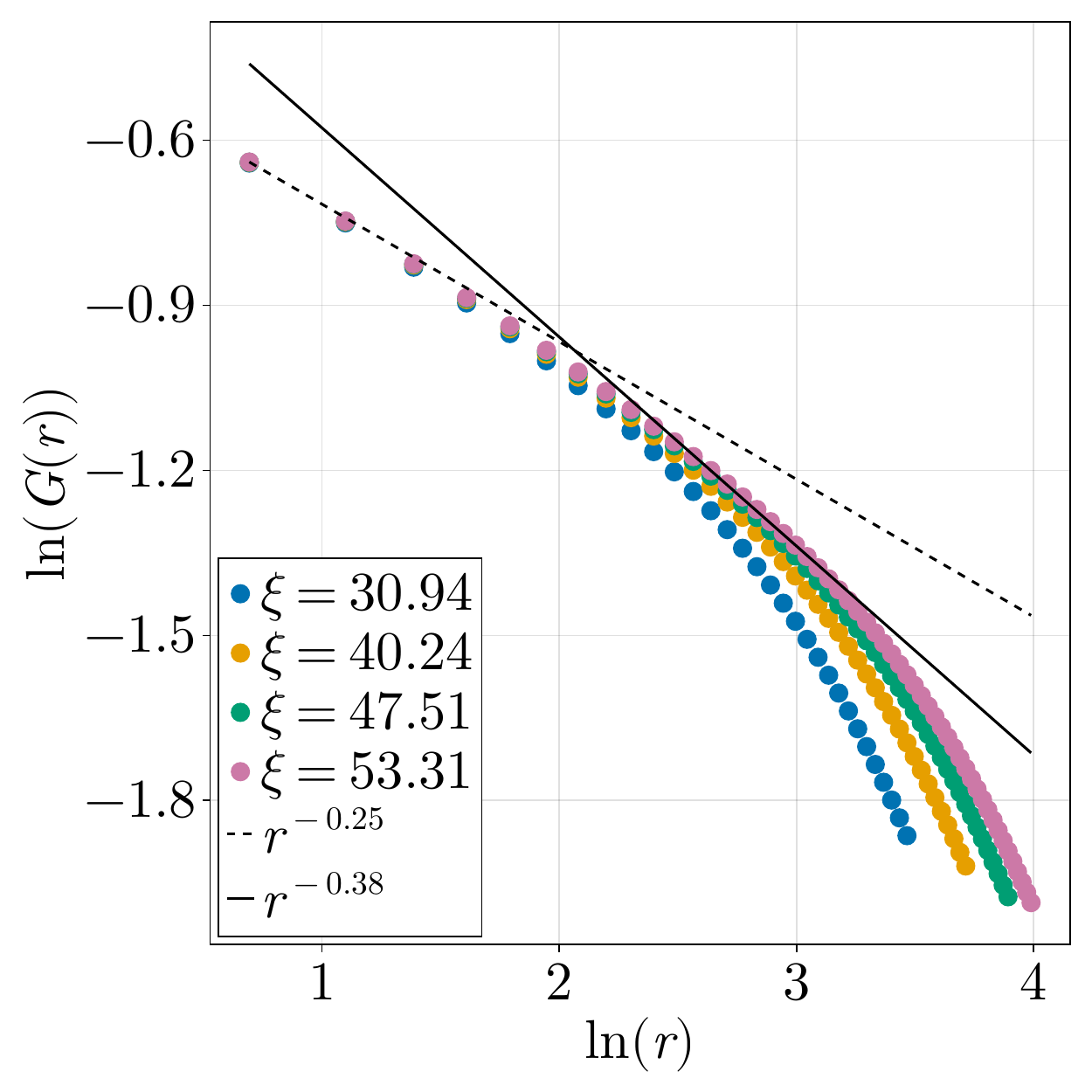} 
    \includegraphics[scale=0.33]{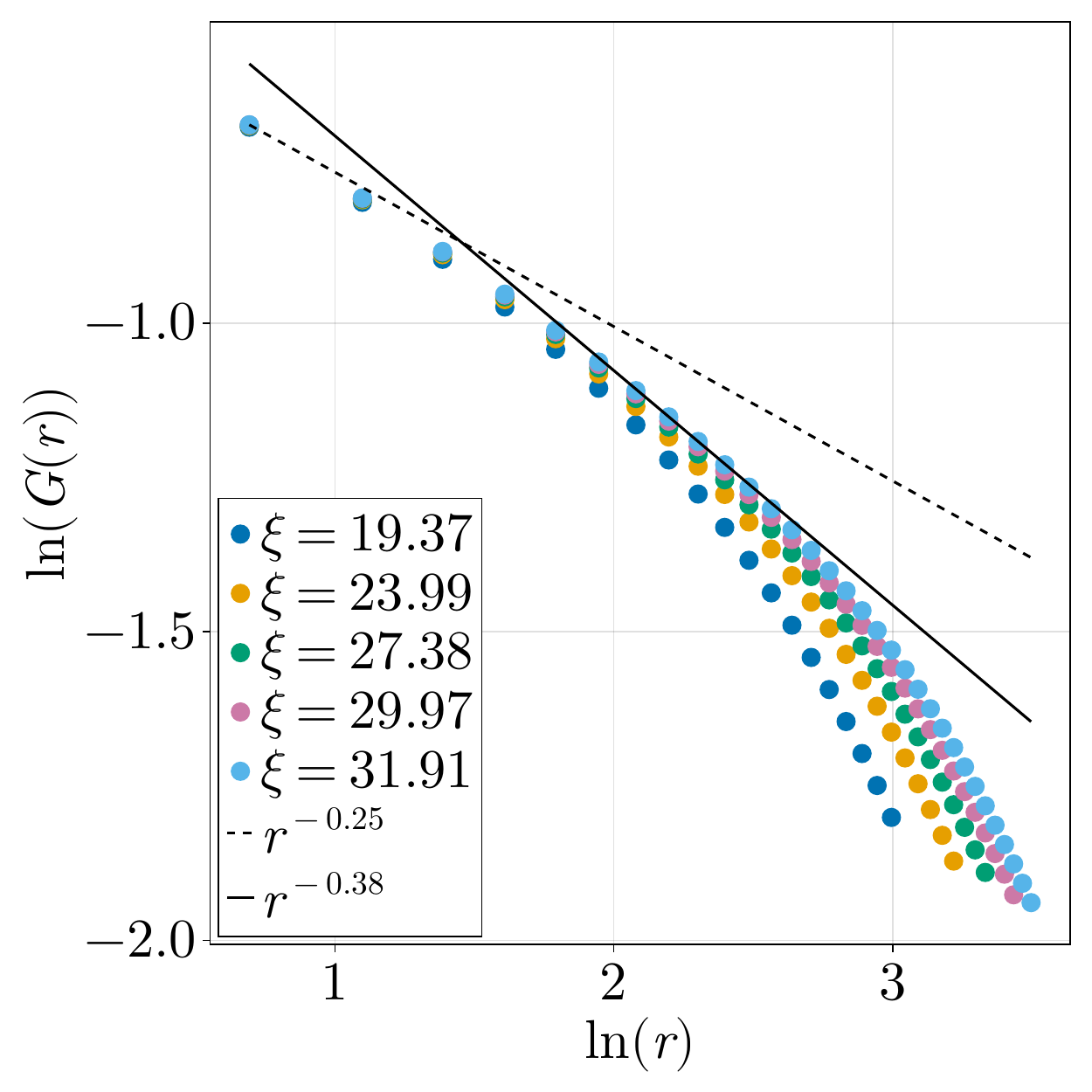} 
    \caption{Left: The correlation functions for different correlation lengths with $D = 80$ and $\Delta\tau = 0.05$. The data sets were obtained by taking $J_n$ and $h_n$ uniformly distributed between $[0.7,1.3]$ with $N_D=9$. The black line represents the algebraic decay of the infinite-randomness fixed point and the dashed line represents the algebraic decay of the clean critical point. Right:  The correlation functions for different correlation lengths with $D = 80$ and $\Delta\tau = 0.05$. The data sets were obtained by taking $J_n$ and $h_n$ uniformly distributed between $[0.5,1.5]$ with $N_D=9$.} \label{fig:correlations}
\end{figure}
From the MPO representation of $\rho$ we obtain the transfer matrix
\begin{equation}
T_{\text{tr}} = \sum_{R}\sum_\sigma P(R)A^{\sigma,\sigma}_R\,,
\end{equation}
whose spectrum determines the decay of disorder-averaged correlation functions. In particular, $\xi = \left[ -\ln(\lambda_2/\lambda_1)\right]^{-1}$, with $\lambda_1$ and $\lambda_2$ respectively the largest and second largest eigenvalue of $T_{\text{tr}}$, is the largest length scale encoded in the MPO. A characteristic property of the infinite-randomness fixed point is that the dynamical exponent is effectively infinite, leading to \emph{activated dynamical scaling}: $\xi \sim (\ln \beta )^2$~\cite{Fisher1992}.  In the left panel of Fig.~\ref{fig:corr_length} we show the correlation length obtained with $D=80$ as a function of $(\ln\beta)^2$, for three different values of $\Delta \tau$. The three curves do not show any significant deviations, indicating that errors due to the Suzuki-Trotter decomposition of $e^{-\Delta\tau H}$ can be neglected. In the middle panel of Fig.~\ref{fig:corr_length} we plot $\xi$ obtained with $\Delta \tau=0.05$ and different bond dimensions as a function of $(\ln\beta)^2$. As expected, at smaller bond dimension $\xi$ deviates from the theoretical curve at smaller $\ln\beta$. We also see that the results are approximately converged for $D=100$ (as the data obtained with $D=80$ and $D=100$ are indistinguishable up to $\beta = 30$), from which we conclude that finite bond dimension-errors have become negligible in this temperature range. 

In the right panel of Fig.~\ref{fig:corr_length} we show the correlation length as a function of $(\ln\beta)^2$ for different numbers of disorder values $N_D = N^2$. We use the same distributions for both $J_n$ and $h_n$, corresponding to a uniform distribution with values $\{0.7, 1.3\}$, $\{0.7, 1.0, 1.3\}$ and $\{0.7, 0.9, 1.1, 1.3\}$ for $N=2$, $N=3$ and $N=4$ respectively. The slope of the correlation length as a function of $(\ln\beta)^2$ should scale with the inverse of the variance of $\ln(J_n)$ and $\ln(h_n)$~\cite{Fisher1992} \footnote{Note that in Ref. ~\cite{Fisher1992} it is stated that all lengths are measured in units of $l_V = \frac{2}{V_I}$. Giving the dependence of $\xi \sim \frac{(\ln\beta)^2}{\text{Var}(\ln h_n)+\text{Var}(\ln J_n)}$}. For the uniform distributions used here these variances are $\text{Var}_{N=2} \approx 0.192$, $\text{Var}_{N=3} \approx 0.097$ and  $\text{Var}_{N=4} \approx 0.0712$. In agreement with theory, in the right panel of Fig.~\ref{fig:corr_length} we indeed see that $\xi/(\ln \beta)^2$ grows when $N$ increases and hence the variance becomes smaller.

The left panel of Fig.~\ref{fig:correlations} shows the disorder-averaged spin correlation function $G(r)$ (i.e. the $\sigma^z-\sigma^z$ correlation function) on a log-log plot for different values of $\xi$. These results were obtained for $N_D=9$ at different temperatures, using $\Delta \tau = 0.05$ and $D=80$. We see that for larger $\xi$ increasingly more data points lie on the black straight line with a negative slope of $2-\phi \approx 0.38$ (with $\phi$ the golden ratio), which is the exact exponent of the average spin correlation function at the infinite randomness fixed point~\cite{Fisher1992,Fisher1995}. In the right panel of Fig.~\ref{fig:correlations} we show the correlation functions obtained with a broader uniform distribution with values $[0.5, 1.0, 1.5]$ for both $J_n$ and $h_n$. We observe that the crossover of $G(r)$ from clean Ising behavior to infinite randomness behavior now occurs at a smaller length scale, as expected from the increase in disorder strength. 

 Additional numerical results are presented in the appendix, such as e.g. $\xi$ as a function of $\delta$, and $\xi(T)$ on the ordered side of the transition, which we confirm grows algebraically with varying exponent, i.e. $\xi \sim 1/T^{\alpha(\delta)}$~\cite{Fisher1995} (as opposed to $\xi \sim e^{\alpha'(\delta)/T}$ in the clean case). The appendix also contains further information on the performance of our algorithm.

\section{Distribution of correlation lengths for the random transverse field Ising model}

\begin{figure}[!htb]
\begin{center}
    \includegraphics[scale=0.30]{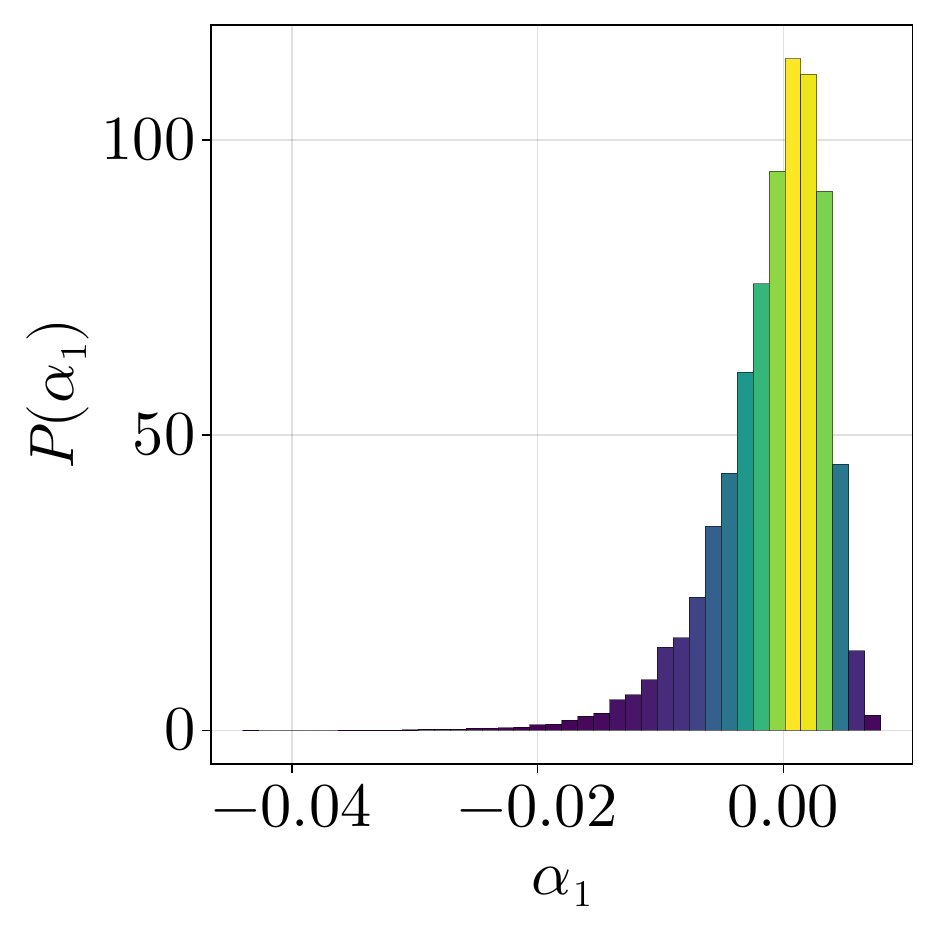} 
    \includegraphics[scale=0.30]{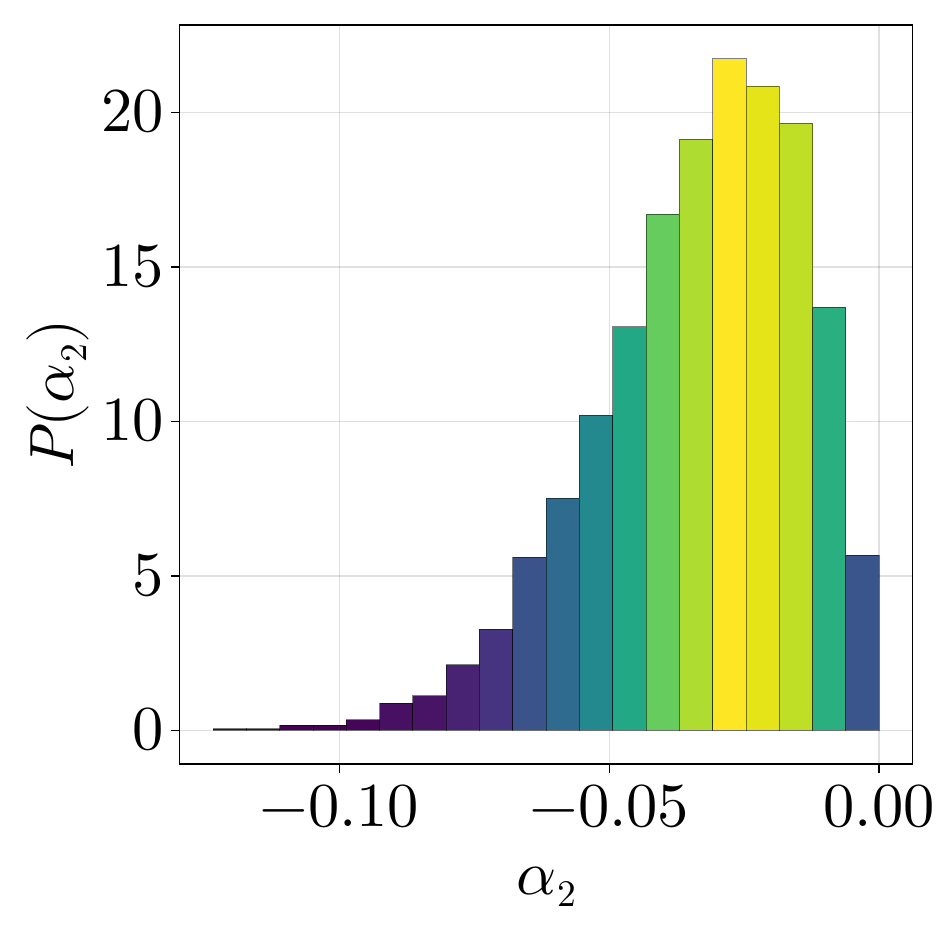} 
    \includegraphics[scale=0.30]{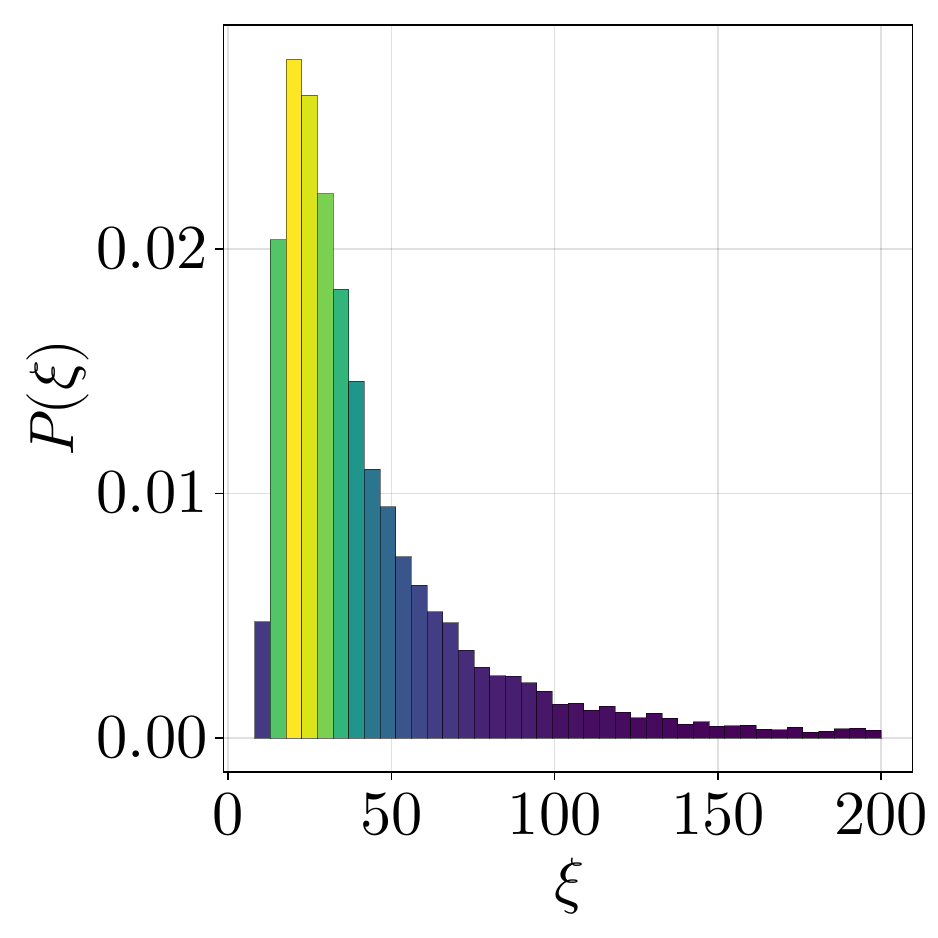} 
    \end{center}
    \caption{Left: Distribution of the first lyapunov exponent.  Middle: Distribution of the second lyapunov exponent. Right: Distribution of the correlation length. All the data was gathered by sampling the MPO for $\beta = 40$, $D=80$ and $\Delta\tau = 0.05$. To compute the lyapunov exponents we used $L=100$ and 20000 samples.} \label{fig:dists}
\end{figure}

In the previous section we extracted the decay of average correlation functions from the transfer matrix spectrum of the average Gibbs state $\rho(\beta)$, which we obtained by tracing out the disorder qudits in $\tilde{\rho}(\beta)$. In this section, however, we consider the transfer matrices of $\tilde{\rho}(\beta)$ without tracing out the disorder qudits:  
\begin{equation}
T_{R} = \sum_\sigma A^{\sigma,\sigma}_R
\end{equation}
For each disorder value $R$, $T_R$ is a $D\times D$ matrix. The different $T_R$ thus provide us with an ensemble of transfer matrices, to which we can apply techniques from random matrix theory. In particular, we can obtain information about the distribution of the correlation length by computing the largest and second largest Lyapunov exponent, $\alpha_1$ and $\alpha_2$. For this we generate two orthogonal random vectors $v_1^{(0)}$ and $v_2^{(0)}$, draw a random matrix $T_{R_1}$ from the ensemble and apply this to obtain two new vectors $v_1^{(1)} = T_{R_1}v_1^{(0)}$ and $v_2^{(1)} = T_{R_1} v_2^{(0)}$. We then re-orthogonalize both vectors and repeat the process by generating a new random matrix $T_{R_2}$ and applying it to the new orthogonal vectors. This process is repeated $L$ times. The first and second Lyapunov exponents are then obtained as $\alpha_i = \frac{1}{L}\log||{v^{(L)}_i}||$. For a given sample of random matrices and large $L$ this procedure is guaranteed to converge by Oseledets' theorem~\cite{oseledets1968multiplicative}. We repeat this procedure for many different samples of the $L$ random matrices to arrive at an estimate of the probability distribution for the Lyapunov exponents. The distributions obtained with $L=100$ and $20.000$ samples for the random transverse field Ising chain are shown in the left and middle panels of Fig. \ref{fig:dists}. These results were obtained at the critical point, i.e. with $\delta =0$, and at an inverse temperature $\beta = 40$.

\begin{figure}[!htb]
\begin{center}
    \includegraphics[scale=0.38]{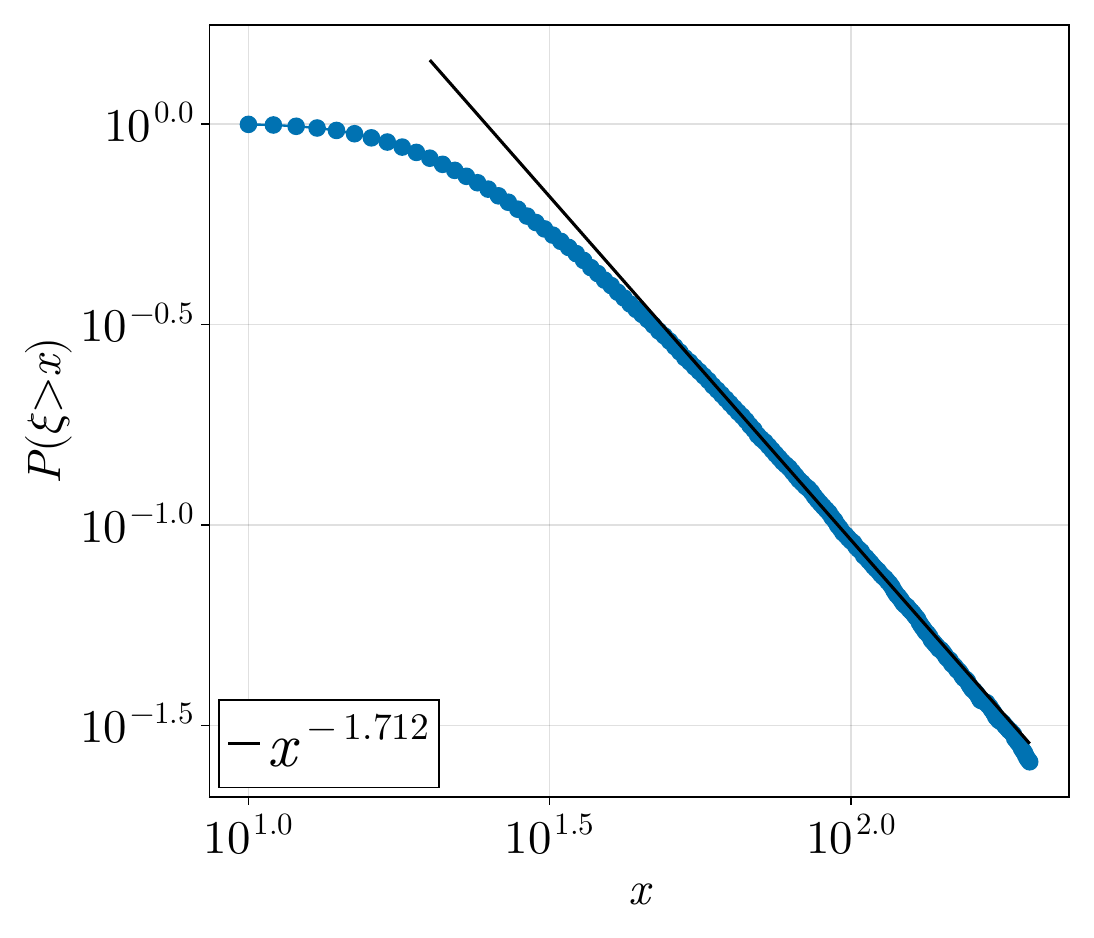} 
    \end{center}
    \caption{The probability mass of the tail of the distribution of correlation lengths. The data was gathered by sampling the MPO for $\beta = 40$, $D=80$ and $\Delta\tau = 0.05$. To compute the lyapunov exponents we used $L=100$ and 20000 samples. } \label{fig:tail}
\end{figure}

In a given disorder configuration, the decay of correlations is determined by random products of the transfer matrices $T_R$. We can therefore obtain the correlation length from the Lyapunov exponents as $\xi = (\alpha_1-\alpha_2)^{-1}$. In the right panel of Fig. \ref{fig:dists} we show the distribution of the correlation length, obtained by computing $\xi$ for each of the different samples. We see that the correlation length distribution has a heavy tail towards large values -- a characteristic feature of the rare-region physics underlying the infinite-randomness physics in the random transverse field Ising chain. Because of this tail, the typical correlation length differs significantly from the average correlation length. If we define the typical correlation length via the \emph{geometric mean} of the correlation function, then we can extract it from our distribution of correlation lengths via the formula
\begin{equation}
\xi_{\text{typ}}^{-1} = \langle \xi^{-1}\rangle\,,
\end{equation}
where the average is over the different samples. The tail of the distribution, which contributes significantly to the average correlation length, has little effect of the typical correlation length. As a result, the typical correlation length $\xi_{\text{typ}}\approx35.4$ is smaller than the average correlation length $\xi_{\text{av}} \approx 47.5$, in agreement with theory~\cite{Fisher1995}. Unfortunately, we are not aware of any theoretical predictions for the temperature dependence of the typical correlation length, which prevents us from explicitly  checking our numerical values for $\xi_{\text{typ}}$. 

 To gain further information on the correlation length distribution produced by our simulations we consider the probability $P(\xi > x)$, which we plot on a log-log scale in Fig. \ref{fig:tail}. We see that the tail of this distribution decays algebraically. From a linear fit to the log-log plot we find the asymptotic behaviour $P(\xi > x) \sim x^{-1.71}$ at $\beta = 40$.

\section{Conclusion}
Our work provides a proof-of-principle that the mixture of Gibbs states with different disorder configurations can be efficiently represented as a single translationally-invariant MPO. The specific algorithm developed here for constructing this MPO allows for many immediate improvements. For example, one could use the better-motivated density matrix truncation (DMT) method for the MPO representation of density matrices~\cite{White2018}. While we found that in the temperature range studied in this work the errors introduced by the Suzuki-Trotter decomposition were not dominant, we still expect it to become important at higher $\beta$. To mitigate these errors one could use the recently developed MPO representation of the cluster expansion of $e^{-
\Delta \tau H}$~\cite{VanderstraetenCluster_2021,VanheckeCluster_2021}, which produces errors that are systematically higher order in $\Delta \tau$. In principle, the cluster expansion method can also be used to reduce the number of imaginary-time steps by increasing $\Delta \tau$ without significant loss of accuracy, but it is unclear whether the inversion of $\Lambda$, crucial for maintaining the correct normalization, can still be done efficiently in that case.

An interesting question for future work is whether a variational formulation of our algorithm could be developed to directly access disorder-averaged ground-state properties. 

While ancilla disorder qudits have previously been used in Ref.~\cite{Paredes2005} for studying average ground-state properties in random spin chains, the approach there relies on adiabatic time-evolution; it would therefore be desirable to obtain the ground state more directly via a variational approach.
It would also be interesting to explore the possibility of combining our approach with the numerical strong-disorder renormalization group method~\cite{igloi2005strong,petHo2025random} for ground state calculations.

\section*{Open-Source Code}
The code used for this work is open-source and is being actively developed and maintained at \url{https://github.com/KVerv/DisorderKit.jl}.

\section*{Acknowledgements}
N.B. acknowledges stimulating discussions with Nicolas Laflorencie, Bram Vanhecke, Jeanne Colbois and Loic Herviou.

% TODO: include author contributions
% \paragraph{Author contributions}
% This is optional. If desired, contributions should be succinctly described in a single short paragraph, using author initials.

% TODO: include funding information
\paragraph{Funding information}
This research was supported by the European Research Council under the European Union Horizon 2020 Research and Innovation Programme via Grant Agreement No. 101076597-SIESS (N.B.). W.T. is funded by the Research Foundation Flanders postdoctoral Fellowship 12AA225N.

\begin{appendix}

\section{Additional details of the VOMPS algorithm for MPO inversion}

An essential part of the algorithm proposed in the main text is the inversion of the MPO $\Lambda$. To find the inverse of an MPO, we variationally optimize the fidelity between the product of the MPO and its candidate inverse and the identity operator. 
To perform this optimization, we make use of the tangent-space method~\cite{haegeman-post-2013,Vanderstraeten2019,Vanhecke2021}. 
More specifically, using the tangent-space method we derive a set of fixed-point equations, and the optimal solution for the MPO inverse can be obtained by iterating these fixed-point equations.

To start, we represent our ansatz for the inverse $\Lambda^{-1}$ as an MPO $\Lambda^{-1}_M$ of bond dimension $\chi$. 
In order to employ the tangent-space techniques, we reformulate the problem into an MPS problem by bending one of the vertical legs of the MPO downward, as shown in Eq.~\eqref{eq:VOMPS1}.
This setup is very similar to that of the VOMPS algorithm~\cite{Vanhecke2021}, which approximates a product of an MPS and an MPO by another MPS with a smaller bond dimension. 
In our case, we look for an MPS such that the product of the MPS and an MPO  approximates a given MPS. 
\begin{equation}
\label{eq:VOMPS1}
\begin{tikzpicture}[scale=1, baseline={height("$=$")}] %[x={10.0pt},y={10.0pt}]
    \node[draw, thick, shape=circle] (v0) at (0,0) {};
    \node[draw, thick, shape=circle] (v1) at (1,0) {};
    \node[draw, thick, shape=circle] (v2) at (2,0) {};
    \node[draw, thick, shape=circle, fill=black!20] (w0) at (0,0.5) {};
    \node[draw, thick, shape=circle, fill=black!20] (w1) at (1,0.5) {};
    \node[draw, thick, shape=circle, fill=black!20] (w2) at (2,0.5) {};
    \draw [thick] 
    (-0.5,0) -- (v0)
    (v0) -- (v1)
    (v0) -- (v1)
    (v1) -- (v2)
    (v2) -- (2.5,0)
    (-0.5,0.5) -- (w0)
    (w0) -- (w1)
    (w1) -- (w2)
    (w2) -- (2.5,0.5);
    \draw [thick, dashed, red]
    (0, -0.5) -- (v0)
    (0, 1) -- (w0)
    (v0) -- (w0)
    (1, -0.5) -- (v1)
    (1, 1) -- (w1)
    (v1) -- (w1)
    (2, -0.5) -- (v2)
    (w2) -- (2, 1)
    (v2) -- (w2);
    \draw  [thick, dashed, red]  (0.25,1) arc (0:180:0.125)
    (0.25,1) -- (0.25, -0.5)
    (1.25,1) arc (0:180:0.125)
    (1.25,1) -- (1.25, -0.5)
    (2.25,1) arc (0:180:0.125)
    (2.25,1) -- (2.25, -0.5);
\end{tikzpicture}
\quad
\approx
\quad
\begin{tikzpicture}[scale=1, baseline={height("$=$")}] %[x={10.0pt},y={10.0pt}]
    \draw [thick, dashed, red]
    (0, -0.5) -- (0, 1)
    (1, -0.5) -- (1, 1)
    (2, -0.5) -- (2, 1);
    \draw  [thick, dashed, red]  (0.25,1) arc (0:180:0.125)
    (0.25,1) -- (0.25, -0.5)
    (1.25,1) arc (0:180:0.125)
    (1.25,1) -- (1.25, -0.5)
    (2.25,1) arc (0:180:0.125)
    (2.25,1) -- (2.25, -0.5);
\end{tikzpicture}
\end{equation}
For later convenience, we will denote the MPS with filled circles and bent dashed line and the MPO with white circles and bent dashed line in Eq.~\eqref{eq:VOMPS1} as
\begin{equation}
\label{eq:MPSnotation}
\hspace{2cm}
    \begin{tikzpicture}[scale=1, baseline={(current bounding box.center)}] %[x={10.0pt},y={10.0pt}]
    \node[draw, thick, shape=circle, fill=black!20] (w0) at (0,0) {};
    \draw [thick] 
    (-0.5,0) -- (w0)
    (w0) -- (0.5,0);
    \draw [thick, dashed, red]
    (0, 0.5) -- (w0)
    (w0) -- (0,-0.5);
    \draw  [thick, dashed, red]  (0.25,0.5) arc (0:180:0.125)
    (0.25,0.5) -- (0.25, -0.5);
\end{tikzpicture}
\rightarrow\begin{tikzpicture}[scale=1., baseline={(current bounding box.center)}]

    \def\l{.7};

    \clip (-1.1*\l,-2.6*\l) rectangle (6.2*\l,2.6*\l); % \clip
    \node[circle, draw, inner sep=1pt] (A1bot) at (0,0) {$A$};
    \draw [dashed, red] (A1bot) --++ (0,-1.*\l);
    \draw (A1bot) --++ (1.*\l,0);
    \draw (A1bot) --++ (-1.*\l,0);
\end{tikzpicture}\hspace{-2cm}
    \begin{tikzpicture}[scale=1, baseline={(current bounding box.center)}] %[x={10.0pt},y={10.0pt}]
    \node[draw, thick, shape=circle] (w0) at (0,0) {};
    \draw [thick] 
    (-0.5,0) -- (w0)
    (w0) -- (0.5,0);
    \draw [thick, dashed, red]
    (0, 0.5) -- (w0)
    (w0) -- (0,-0.5);
    \draw  [thick, dashed, red]
    (0.25,0.5) -- (0.25, -0.5);
\end{tikzpicture}
\rightarrow \begin{tikzpicture}[scale=1., baseline={(current bounding box.center)}]

    \def\l{.7};

    \clip (-1.1*\l,-2.6*\l) rectangle (6.2*\l,2.6*\l); % \clip
    \node[circle, draw, inner sep=1pt] (A1bot) at (0,0) {$O$};
    \draw [dashed, red] (A1bot) --++ (0,-1.*\l);
    \draw [dashed, red] (A1bot) --++ (0,1.*\l);
    \draw (A1bot) --++ (1.*\l,0);
    \draw (A1bot) --++ (-1.*\l,0);
\end{tikzpicture}
\end{equation}
The cost function for the variational optimization corresponds to the fidelity between the two states in Eq.~\eqref{eq:VOMPS1}, which is expressed as  
\begin{equation}
    \mathcal{L} = \ln{\left(\frac{\langle\psi(\bar{A})|\Lambda^\dagger|\mathds{1}\rangle\langle\mathds{1}|\Lambda|\psi(A)\rangle}{\langle\psi(\bar{A})|\Lambda^\dagger \Lambda|\psi(A)\rangle}\right)},
\end{equation}
where we defined $|\mathds{1}\rangle$ as the state representing the trivial MPS on the right hand side of  Eq.~\eqref{eq:VOMPS1}, and $|\psi(A)\rangle$ is the MPS version of $\Lambda^{-1}_M$ (i.e. the MPO with filled circles and bent dashed line in Eq.~\eqref{eq:VOMPS1} and Eq.~\eqref{eq:MPSnotation}). The gradient of this cost function can be computed by taking the derivative with respect to $\bar{A}$. The optimal solution is then found as the point where the gradient vanishes. Writing out this gradient gives
\begin{equation}
    \frac{\partial \mathcal{L}}{\partial \bar{A}} = \langle \partial_{\bar{A}} \psi(\bar{A})|\left(\Lambda^\dagger|\mathds{1}\rangle-\frac{\langle\psi(\bar{A})|\Lambda^\dagger|\mathds{1}\rangle}{\langle\psi(\bar{A})|\Lambda^\dagger \Lambda|\psi(A)\rangle} \Lambda^\dagger \Lambda|\psi(A)\rangle\right).\label{eq:gradient}
\end{equation}
The optimality condition $\partial \mathcal{L} / \partial \bar{A} = 0$ can be reformulated as 
\begin{equation}
   \mathcal{P}_A \left(\Lambda^\dagger|\mathds{1}\rangle-\frac{\langle\psi(\bar{A})|\Lambda^\dagger|\mathds{1}\rangle}{\langle\psi(\bar{A})|\Lambda^\dagger \Lambda|\psi(A)\rangle} \Lambda^\dagger \Lambda|\psi(A)\rangle\right) = 0,\label{eq:Pgradient}
\end{equation}
where the projector $\mathcal{P}_A$ is the projector that projects a state onto the tangent space of $|\psi(A)\rangle$, whose explicit form is given by 
\begin{equation}
    \mathcal{P}_A = \sum_i     
    \begin{tikzpicture}[scale=1., baseline={(current bounding box.center)}]

    \def\l{.7};

    \clip (-1.1*\l,-2.6*\l) rectangle (6.2*\l,2.6*\l); % \clip

    % Nodes
    \node[circle, draw, inner sep=1pt] (A1top) at (0, \l) {$ \bar{A}_L$}; % \newcommand{}{\scriptstyle}
    \node[circle, draw, inner sep=1pt] (A1bot) at (0,-\l) {$ A_L$};
    \node[] (A2top) at (2*\l, \l) {};
    \node[] (A2bot) at (2*\l,-\l) {};
    \node[circle, draw, inner sep=1pt] (A3top) at (4*\l, \l) {$ \bar{A}_R$};
    \node[circle, draw, inner sep=1pt] (A3bot) at (4*\l,-\l) {$ A_R$};

    % Horizontal lines
    \draw (A1top) --++ (-1.*\l,0);
    \draw (A1bot) --++ (-1.*\l,0);
    \draw (A3top) --++ (1.*\l,0);
    \draw (A3bot) --++ (1.*\l,0);

    % Vertical lines
    \draw [dashed, red] (A2top) --++ (0,1.*\l);
    \draw [dashed, red] (A2bot) --++ (0,-1.*\l);
    \draw [dashed, red] (A1top) --++ (0,1.*\l);
    \draw [dashed, red] (A1bot) --++ (0,-1.*\l);
    \draw [dashed, red](A3top) --++ (0,1.*\l);
    \draw [dashed, red] (A3bot) --++ (0,-1.*\l);
    \draw [dashed, red] (A2bot.south) -- (A2top.north);

    % Arcs
    \draw (A1top) to[out=0,in=0] (A1bot);
    \draw (A3top) to[out=180,in=180] (A3bot);

    % Site labels
    \node at (0,2.3*\l) {$ i-1$};
    \node at (2*\l,2.3*\l) {$ i$};
    \node at (4*\l,2.3*\l) {$ i+1$};
\end{tikzpicture}
    \hspace{-0.5cm}-\hspace{0.25cm}
    \begin{tikzpicture}[scale=1., baseline={(current bounding box.center)}]

    \def\l{.7};

    \clip (-1.1*\l,-2.6*\l) rectangle (6.2*\l,2.6*\l); % \clip

    % Nodes
    \node[circle, draw, inner sep=1pt] (A1top) at (0, \l) {$ \bar{A}_L$}; % \newcommand{}{\scriptstyle}
    \node[circle, draw, inner sep=1pt] (A1bot) at (0,-\l) {$ A_L$};
    \node[circle, draw, inner sep=1pt] (A2top) at (2*\l, \l) {$\bar{A}_L$};
    \node[circle, draw, inner sep=1pt] (A2bot) at (2*\l,-\l) {$ A_L$};
    \node[circle, draw, inner sep=1pt] (A3top) at (5*\l, \l) {$ \bar{A}_R$};
    \node[circle, draw, inner sep=1pt] (A3bot) at (5*\l,-\l) {$ A_R$};

    % Horizontal lines
    \draw (A2top) -- (A1top);
    \draw (A2bot) -- (A1bot);
    \draw (A1top) --++ (-1.*\l,0);
    \draw (A1bot) --++ (-1.*\l,0);
    \draw (A3top) --++ (1.*\l,0);
    \draw (A3bot) --++ (1.*\l,0);

    % Vertical lines
    \draw [dashed, red] (A2top) --++ (0,1.*\l);
    \draw [dashed, red] (A2bot) --++ (0,-1.*\l);
    \draw [dashed, red] (A1top) --++ (0,1.*\l);
    \draw [dashed, red] (A1bot) --++ (0,-1.*\l);
    \draw [dashed, red] (A3top) --++ (0,1.*\l);
    \draw [dashed, red] (A3bot) --++ (0,-1.*\l);

    % Arcs
    \draw (A2top) to[out=0,in=0] (A2bot);
    \draw (A3top) to[out=180,in=180] (A3bot);

    % Site labels
    \node at (0,2.3*\l) {$ i-1$};
    \node at (2*\l,2.3*\l) {$ i$};
    \node at (5*\l,2.3*\l) {$ i+1$};
\end{tikzpicture}
\end{equation}
The condition \eqref{eq:Pgradient} allows us to derive a set of consistency equation that the optimal solution should satisfy. Denoting the local tensor of the MPO $\Lambda$ as $O$, Eq.\eqref{eq:Pgradient} leaves us with the following consistency equations for the optimal solution:
\begin{equation}
    \label{eq:VOMPS_AC}
\begin{tikzpicture}[scale=1, baseline={height("$=$")}] %[x={10.0pt},y={10.0pt}]
    \node[draw, thick, shape=circle] (v0) at (0,0) {$\rho_L$};
    \node[draw, thick, shape=circle] (v1) at (1.5,0) {$O^\dagger$};
    \node[draw, thick, shape=circle] (v2) at (3,0) {$\rho_R$};
    \node[draw, thick, shape=circle] (id1) at (1.5,1.5) {$\mathds{1}$};
    \draw [thick]
    (v0) -- (v1)
    (v0) -- (v1)
    (v1) -- (v2)
    (v0.south) arc (0:-90:0.5)
    (v2.south) arc (180:270:0.5);
    \draw [thick, dashed, red]
    (1.5, -1) -- (v1);
    \draw  [thick, dashed, red]
    (v1.north) -- (id1);
\end{tikzpicture}
= \frac{N_\rho}{N_E}
\begin{tikzpicture}[scale=1, baseline={height("$=$")}] %[x={10.0pt},y={10.0pt}]
    \node[draw, thick, shape=rectangle, minimum height=10em] (v0) at (0,0.8) {$E_L$};
    \node[draw, thick, shape=circle] (v1) at (1.5,0) {$O^\dagger$};
    \node[draw, thick, shape=circle] (v3) at (1.5,1.15) {$O$};
    \node[draw, thick, shape=rectangle, minimum height=10em] (v2) at (3,0.8) {$E_R$};
    \node[draw, thick, shape=circle, fill=black!20] (w0) at (1.5,2.25) {$A_C$};
    \draw [thick]
    (v1.west) -- (v0.east |- v1.east)
    (v3.west) -- (v0.east |- v3.east)
    (w0.west) -- (v0.east |- w0.east)
    (v1.east) -- (v2.west |- v1.east)
    (v3.east) -- (v2.west |- v3.east)
    (w0.east) -- (v2.west |- w0.east)
    (v0.south) arc (0:-90:0.5)
    (v2.south) arc (180:270:0.5);
    \draw [thick, dashed, red]
    (1.5, -1.5) -- (v1)
    (v1) -- (v3)
    (v3) -- (w0);
\end{tikzpicture},
\end{equation}

\begin{equation}
    \label{eq:VOMPS_C}
\begin{tikzpicture}[scale=1, baseline={height("$=$")}] %[x={10.0pt},y={10.0pt}]
    \node[draw, thick, shape=circle] (v0) at (0,0) {$\rho_L$};
    \node[draw, thick, shape=circle] (v1) at (1,0) {$\rho_R$};
    \draw [thick]
    (v0) -- (v1)
    (v0) -- (v1)
    (v0.south) arc (0:-90:0.5)
    (v1.south) arc (180:270:0.5);
\end{tikzpicture}
= \frac{N_\rho}{N_E}
\begin{tikzpicture}[scale=1, baseline={height("$=$")}] %[x={10.0pt},y={10.0pt}]
    \node[draw, thick, shape=rectangle, minimum height=10em] (v0) at (0,0.8) {$E_L$};
    \node[draw, thick, shape=rectangle, minimum height=10em] (v2) at (3,0.8) {$E_R$};
    \node[draw, thick, shape=circle, fill=black!20] (w0) at (1.5,2.25) {$C$};
    \node[] (w1) at (1.5,-0.25) {};
    \draw [thick]
    (w0.west) -- (v0.east |- w0.east)
    (w0.east) -- (v2.west |- w0.east)
    (v0.east) -- (v2.west)
    (v0.east |- w1) -- (v2.west |- w1)
    (v0.south) arc (0:-90:0.5)
    (v2.south) arc (180:270:0.5);
\end{tikzpicture}.
\end{equation}
In these equations we have defined:
\begin{equation}
\label{eq:N_E_and_N_rho}
N_E =
\begin{tikzpicture}[scale=1, baseline={height("$=$")}] %[x={10.0pt},y={10.0pt}]
    \node[draw, thick, shape=circle] (v1) at (1.5,0) {$O^\dagger$};
    \node[draw, thick, shape=circle] (v3) at (1.5,1.15) {$O$};
    \node[draw, thick, shape=rectangle, minimum height=12em] (v2) at (3,0.5) {$E_R$};
    \node[draw, thick, shape=rectangle, minimum height=12em] (v0) at (0,0.5) {$E_L$};
    \node[draw, thick, shape=circle, fill=black!20] (w0) at (1.5,2.25) {$A_C$};
    \node[draw, thick, shape=circle, fill=black!20] (w1) at (1.5,-1.25) {$\bar{A}_C$};
    \draw [thick]
    (v1.east) -- (v2.west |- v1.east)
    (v3.east) -- (v2.west |- v3.east)
    (w0.east) -- (v2.west |- w0.east)
    (w1.east) -- (v2.west |- w1.east)
    (v1.west) -- (v0.east |- v1.east)
    (v3.west) -- (v0.east |- v3.east)
    (w0.west) -- (v0.east |- w0.east)
    (w1.west) -- (v0.east |- w1.east)
    (w0.west) -- (0.5,2.25)
    (v3.west) -- (0.5,1.15)
    (v1.west) -- (0.5,0)
    (w1.west) -- (0.5,-1.25);
    \draw [thick, dashed, red]
    (w1) -- (v1)
    (v1) -- (v3)
    (v3) -- (w0);
\end{tikzpicture} \; ,
%\end{equation}
%\begin{equation}
%\label{eq:N_rho}
\quad
N_\rho =
\begin{tikzpicture}[scale=1, baseline={height("$=$")}] %[x={10.0pt},y={10.0pt}]
    \node[] (v0) at (0,0.75) {};
    \node[] (v3) at (0,-.75) {};
    \node[draw, thick, shape=circle] (v1) at (1.5,0.75) {$O^\dagger$};
    \node[draw, thick, shape=circle] (v2) at (3,0) {$\rho_R$};
    \node[draw, thick, shape=circle] (v4) at (0,0) {$\rho_L$};
    \node[draw, thick, shape=circle, fill=black!20] (w0) at (1.5,-0.75) {$\bar{A}_C$};
    \node[draw, thick, shape=circle] (id1) at (1.5,2) {$\mathds{1}$};
    \draw [thick]
    (v0) -- (v1)
    (w0) -- (v3)
    (w0.east) -- (v2.south |- w0.east)
    (v2.south) -- (v2.south |- w0.east)
    (v1.east) -- (v2.north |- v1.east)
    (v2.north) -- (v2.north |- v1.east)
    (w0.west) -- (v4.south |- w0.west)
    (v4.south) -- (v4.south |- w0.west)
    (v1.west) -- (v4.north |- v1.west)
    (v4.north) -- (v4.north |- v1.west);
    \draw [thick, dashed, red]
    (v1) -- (w0)
    (v1) -- (id1);
\end{tikzpicture} \; ,
\end{equation}
\begin{equation}
\label{eq:E_env_and_rho_env}
\begin{tikzpicture}[scale=1, baseline={height("$=$")}] %[x={10.0pt},y={10.0pt}]
    \node[draw, thick, shape=circle] (v1) at (1.5,0) {$O^\dagger$};
    \node[draw, thick, shape=circle] (v3) at (1.5,1.15) {$O$};
    \node[draw, thick, shape=rectangle, minimum height=12em] (v2) at (3,0.5) {$E_R$};
    \node[draw, thick, shape=circle, fill=black!20] (w0) at (1.5,2.25) {$A_R$};
    \node[draw, thick, shape=circle, fill=black!20] (w1) at (1.5,-1.25) {$\bar{A}_R$};
    \draw [thick]
    (v1.east) -- (v2.west |- v1.east)
    (v3.east) -- (v2.west |- v3.east)
    (w0.east) -- (v2.west |- w0.east)
    (w1.east) -- (v2.west |- w1.east)
    (w0.west) -- (0.5,2.25)
    (v3.west) -- (0.5,1.15)
    (v1.west) -- (0.5,0)
    (w1.west) -- (0.5,-1.25);
    \draw [thick, dashed, red]
    (w1) -- (v1)
    (v1) -- (v3)
    (v3) -- (w0);
\end{tikzpicture}
= \lambda_E
\begin{tikzpicture}[scale=1, baseline={height("$=$")}] %[x={10.0pt},y={10.0pt}]
    \node[] (v1) at (1.5,0) {};
    \node[] (v3) at (1.5,1.15) {};
    \node[draw, thick, shape=rectangle, minimum height=12em] (v2) at (3,0.5) {$E_R$};
    \node[] (w0) at (1.5,2.25) {};
    \node[] (w1) at (1.5,-1.25) {};
    \draw [thick]
    (v1.east) -- (v2.west |- v1.east)
    (v3.east) -- (v2.west |- v3.east)
    (w0.east) -- (v2.west |- w0.east)
    (w1.east) -- (v2.west |- w1.east);
\end{tikzpicture} \; ,
\quad
%\end{equation}
%\begin{equation}
%\label{eq:rho_env}
\begin{tikzpicture}[scale=1, baseline={height("$=$")}] %[x={10.0pt},y={10.0pt}]
    \node[] (v0) at (0.5,0.75) {};
    \node[] (v3) at (0.5,-.75) {};
    \node[draw, thick, shape=circle] (v1) at (1.5,0.75) {$O^\dagger$};
    \node[draw, thick, shape=circle] (v2) at (3,0) {$\rho_R$};
    \node[draw, thick, shape=circle, fill=black!20] (w0) at (1.5,-0.75) {$\bar{A}_R$};
    \node[draw, thick, shape=circle] (id1) at (1.5,2) {$\mathds{1}$};
    \draw [thick]
    (v0) -- (v1)
    (w0) -- (v3)
    (w0.east) -- (v2.south |- w0.east)
    (v2.south) -- (v2.south |- w0.east)
    (v1.east) -- (v2.north |- v1.east)
    (v2.north) -- (v2.north |- v1.east);
    \draw [thick, dashed, red]
    (v1) -- (w0)
    (v1) -- (id1);
\end{tikzpicture}
= \lambda_\rho
\begin{tikzpicture}[scale=1, baseline={height("$=$")}] %[x={10.0pt},y={10.0pt}]
    \node[] (v1) at (1.75,0.75) {};
    \node[draw, thick, shape=circle] (v2) at (3,0) {$\rho_R$};
    \node[] (w0) at (1.75,-0.75) {};
    \draw [thick]
    (w0.east) -- (v2.south |- w0.east)
    (v2.south) -- (v2.south |- w0.east)
    (v1.east) -- (v2.north |- v1.east)
    (v2.north) -- (v2.north |- v1.east);
\end{tikzpicture} \; ,
\end{equation}
and similarly for $\rho_L$ and $E_L$.

In the following, we provide a detailed description of the algorithm.
\begin{enumerate}
\item First we prepare an initial guess for $\Lambda^{-1}_M$. We then bring it to MPS form $|\psi(A)\rangle$ by bending one of the vertical legs of the MPO downward [c.f. Eq.~\eqref{eq:VOMPS1}], and convert the MPS to the mixed canonical form.
The virtual bond dimension for this MPS is the same as the inverse MPO, while the dimension of its open indices is the square of the original disorder qudit dimension. 
\item We construct the environment tensors $\rho_L$, $\rho_R$, $E_L$, and $E_R$ by solving the eigenvalue equations \eqref{eq:E_env_and_rho_env}. 
\item By solving the linear equations in Eqs.~\eqref{eq:VOMPS_AC} and \eqref{eq:VOMPS_C}, we can obtain a new set of $A_C$ and $C$. 
\item From the new set of $A_C$ and $C$, we can obtain a new set of left- and right-canonical tensors $A_L$ and $A_R$, where we simply adopt the procedure used in the standard variational uniform MPS (VUMPS) algorithm~\cite{zauner-stauber-variational-2018}.
\item If $|\psi(A)\rangle$ is already precisely at the optimal point, then for the updated tensors $A_C$, $C$, $A_L$, and $A_R$, we expect both $\epsilon_L = \| A_C - A_L \cdot C \|$ and $\epsilon_R = \| A_C - C \cdot A_R \|$ to vanish. 
In practice, with our initial guess, this is rarely the case. 
Therefore, we need to return to step (ii), and iterate the process from step (ii) to step (iv) until both $\epsilon_L$ and $\epsilon_R$ go below a chosen tolerance.
\item Finally, after the iteration converges, we convert the optimized MPS $|\psi(A^\star)\rangle$ back into the MPO form, yielding the approximate inverse MPO $\Lambda^{-1}_M$.
\end{enumerate}

The algorithm as explained above assumes a single site unit-cell. It is however straightforward to generalize the algorithm to larger unit-cell structures. In this case one needs to solve Eqs.~\eqref{eq:VOMPS_AC} and \eqref{eq:VOMPS_C} separately for every tensor in the unit-cell. Additional modifications can be made to reduce runtime and memory cost. For example, one can make use of the fact that both $\Lambda$ and $\Lambda^{-1}$ are diagonal, and hence can be represented with three-leg tensors. This reduces the memory cost from a quadratic to a linear dependence on the amount of disorder samples. We refer to the open-source code accompanying this work (\url{https://github.com/KVerv/DisorderKit.jl}) for complete details of the algorithm used in this work.

\begin{figure}[!htb]
    a)
    \includegraphics[scale=0.25]{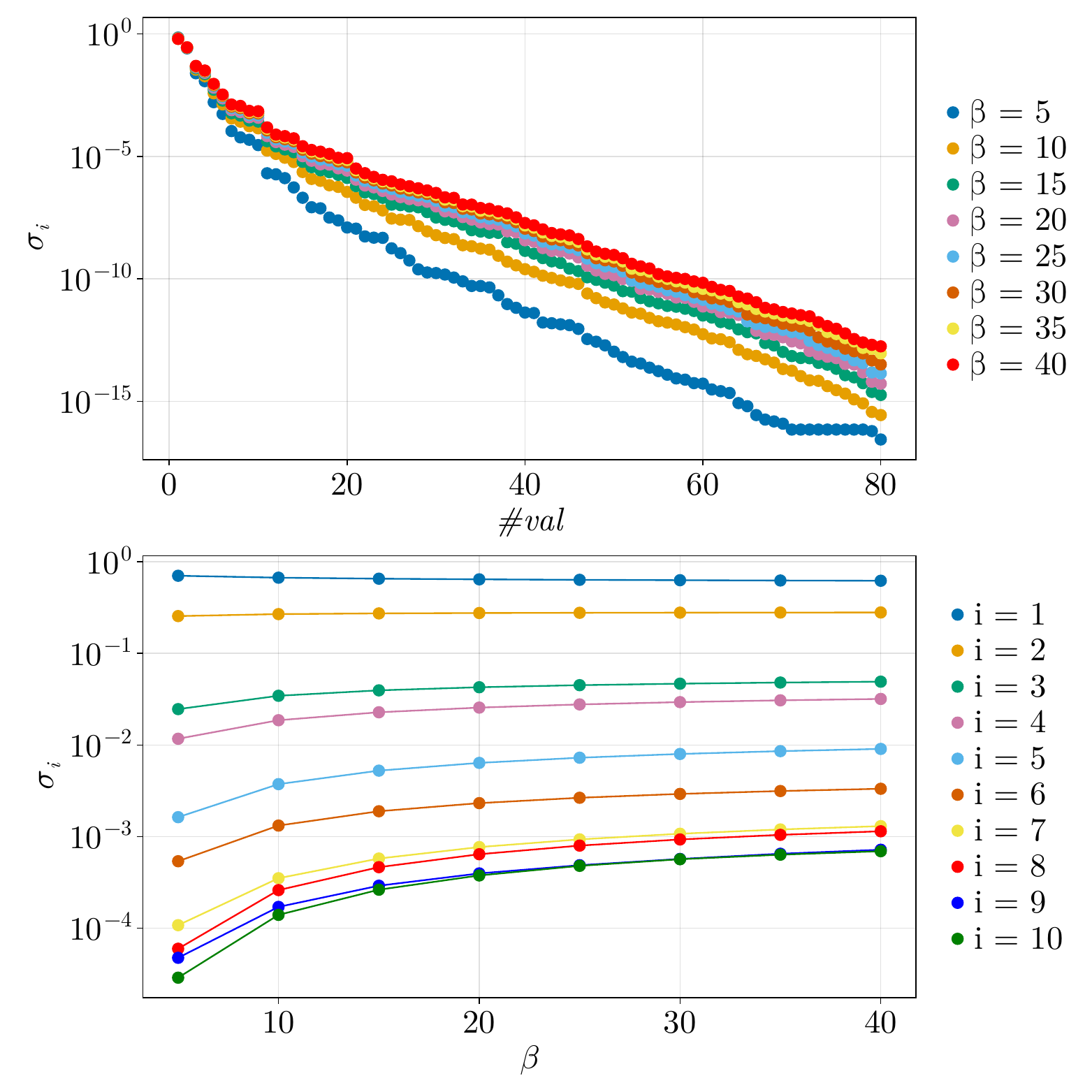}
    \hspace{0.25 cm} b)
    \includegraphics[scale=0.25]{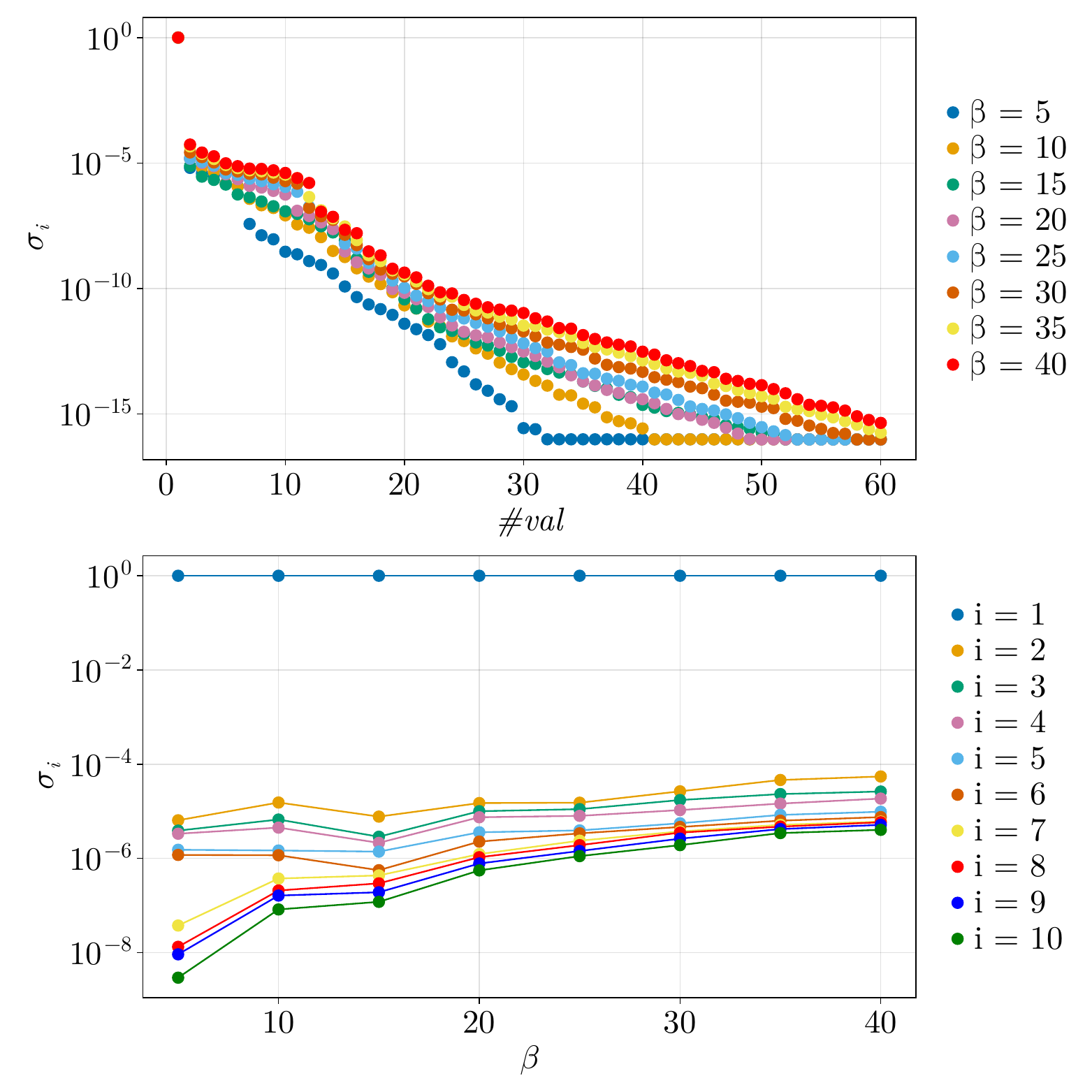} 
    \caption{(a) Operator entanglement spectrum of the disorder averaged density matrix $\rho(\beta)$. (b) Entanglement spectrum of the partition function MPO $\Lambda$. The data was acquired with $D=80$ and $\Delta\tau = 0.05$.} \label{fig:entanglement}
\end{figure}

\section{Performance of the algorithm for the MPO inversion}

 As argued in the main text, obtaining the inverse of $\Lambda$ as a low bond-dimension MPO should be possible as long as the imaginary-time steps are sufficiently small. Here we verify this expectation through several numerical checks.

In Fig.~\ref{fig:entanglement}, we show the operator entanglement spectrum of the disorder averaged density matrix $\rho(\beta)$ (Fig.~\ref{fig:entanglement}(a)) and the partition function MPO $\Lambda$ (Fig.~\ref{fig:entanglement}(b)).
We see that the operator entanglement spectrum of $\Lambda$ decays much faster than that of $\rho(\beta)$, indicating that $\Lambda$ can be truncated to a much lower bond dimension ($D_\Lambda \sim 5$) than $\rho(\beta)$ ($D_\rho \sim 100$).
This is consistent with the MPO $\Lambda$ being close to the identity operator. 

\begin{figure}[!htb]
    \includegraphics[scale=0.25]{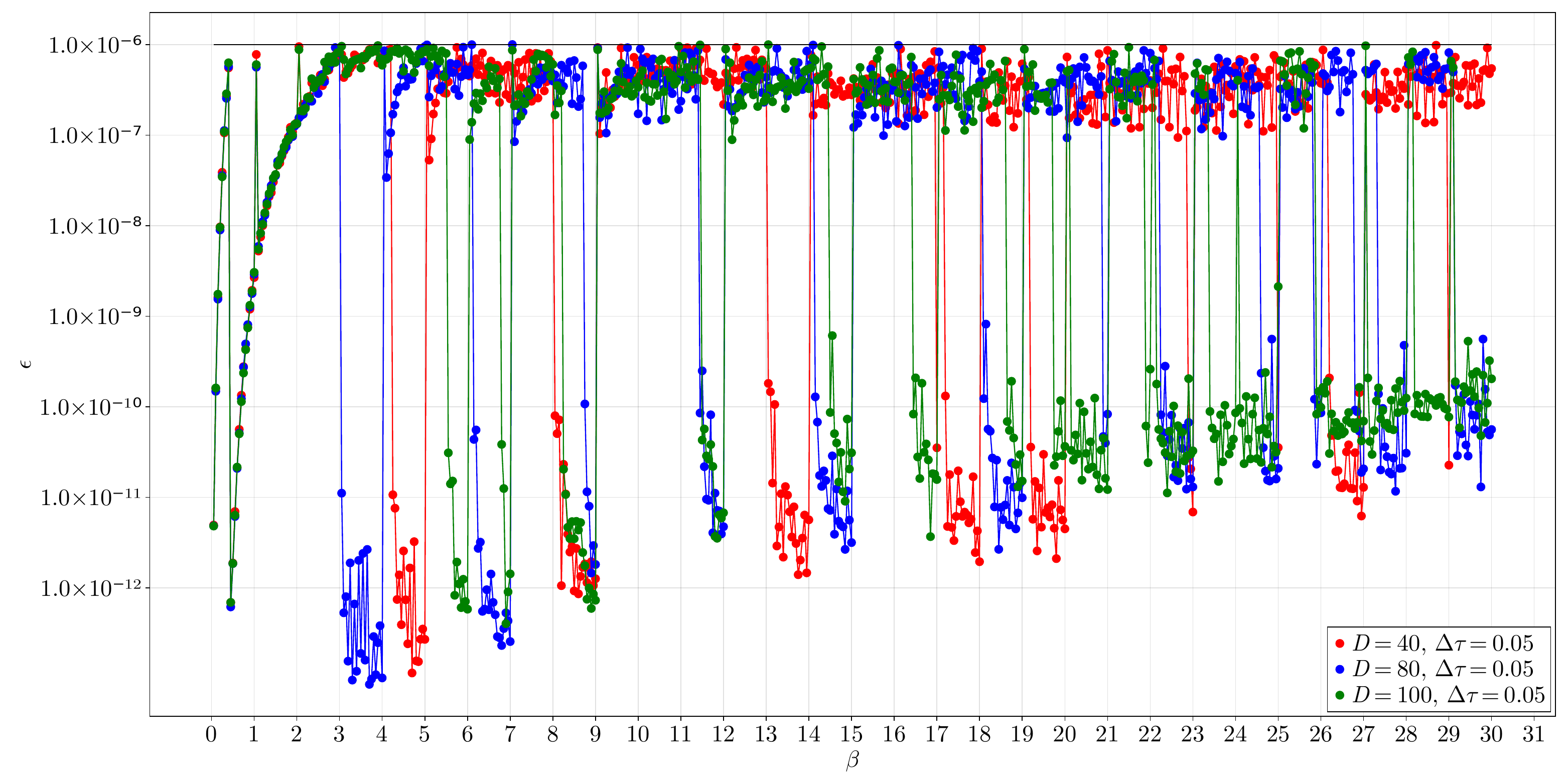} 
    \caption{Error made in the inversion of $\Lambda$ for different bond dimensions $\chi$ of the MPO $\Lambda^{-1}_M$. The error is computed by summing all but the largest Schmidt values of $\Lambda\Lambda^{-1}_M$. The black line denotes the threshold of $\epsilon = 10^{-6}$. } \label{fig:inversion}
\end{figure}

The extremely gapped operator entanglement spectrum of $\Lambda$ also suggests that its inverse should admit a very low bond-dimension MPO approximation.
In our simulations, we first set the bond dimension $\chi$ of $\Lambda^{-1}_M$ to be 1, and then increase $\chi$ when the error exceeds the threshold $\epsilon = 10^{-6}$.
In Fig.~\ref{fig:inversion}, we show the error introduced in the approximate inversion of $\Lambda$ for different inverse temperatures $\beta$.
For a given $\chi$, the inversion becomes less accurate as $\beta$ increases.
When it exceeds the threshold $\epsilon = 10^{-6}$, we increase $\chi$, and the error in the inversion drops again. In Fig.~\ref{fig:inversion} this corresponds to the sudden discontinuous jumps to lower inversion errors .

We also notice that it is not necessary to always keep increasing $\chi$ or keep $\chi$ fixed at a certain relatively large value after it has been increased. 
In our simulations, we try to reset $\chi$ back to 1 when $\beta$ takes integer values.
This corresponds to the upward discontinuous jumps of the inversion error at integer $\beta$ in Fig.~\ref{fig:inversion}.
Although the inversion error increases after resetting $\chi$ to 1, we notice that it often still stays below the threshold $\epsilon = 10^{-6}$.
This allows us to use a relatively small $\chi$ for the inversion in the following imaginary-time steps before the error exceeds the threshold again. 
In our simulations, the bond dimension $\chi$ never exceeds 4, which is consistent with our previous discussion that the MPO $\Lambda$ is close to the identity operator.

\newpage
\section{Additional results for the Random Transverse Field Ising Model}

\begin{figure}[!htb]
\begin{center}
    \includegraphics[scale=0.4]{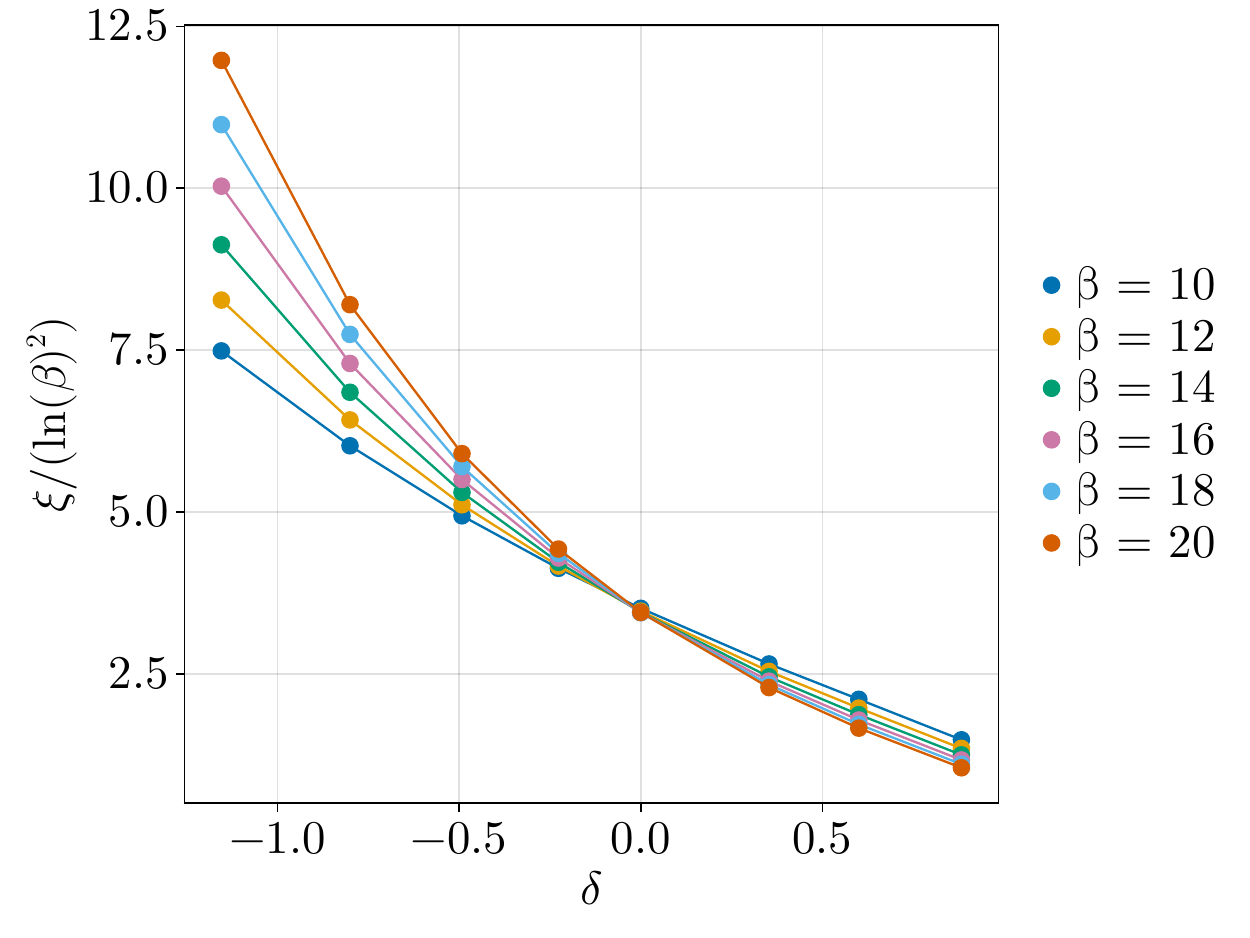} 
    \end{center}
    \caption{Correlation length of the disorder-averaged correlation function as a function of $\delta$ for different temperatures. The different curves cross around $\delta \approx 0$, which corresponds to $h_\text{max}\approx 1.3$.} \label{fig:corr_delta}
\end{figure}

The relevant quantity characterizing the behavior of the RTFIM is 
\begin{equation}
\delta = [\langle\ln h_i\rangle-\langle\ln J_i\rangle]/[\text{Var}(\ln h_i)+\text{Var}(\ln J_i)]    
\end{equation} 
For $\delta>0$, the model is in the paramagnetic phase, whereas the model exhibits long-range order and $\mathbb{Z}_2$ symmetry breaking for $\delta < 0$. 
On the ordered side of the quantum phase transition, the finite-temperature correlation length grows as $\xi \sim 1/T^{\alpha(\delta)}$~\cite{Fisher1995}, i.e. the correlation length diverges as power of $1/T$ with a continuously varying exponent. 
This behavior is different from the clean Ising model, where the correlation length grows as $\xi \sim e^{\alpha'(\delta)/T}$. 
On the disordered side the correlation length is predicted to scale as $\xi \sim (\ln\beta)^2 /[(\delta-\delta_c)^2 (\ln\beta)^2+a]$~\cite{Fisher1995}, with $a$ some constant independent of $\delta$ and $\beta$. For large $\beta$ the correlation length saturates to $\xi\sim1/(\delta-\delta_c)^2$. Motivated by these scaling considerations, we plot $\xi(\delta)/\ln(\beta)^2$ for different $\beta$ as a function of $\delta$ in Fig.~\ref{fig:corr_delta}. The results were obtained by taking $J_n$ and $h_n$ to be uniformly distributed between $[0.7, h_{\text{max}}]$ with a total of $N_D=9$ disorder values. To change $\delta$ we performed simulations with different values of $h_{\text{max}}$ ranging from $1.1$ to $1.7$. The imaginary time step size used for all simulations was $\Delta\tau = 0.05$.

\begin{figure}[!htb]
\begin{center}
    \includegraphics[scale=0.4]{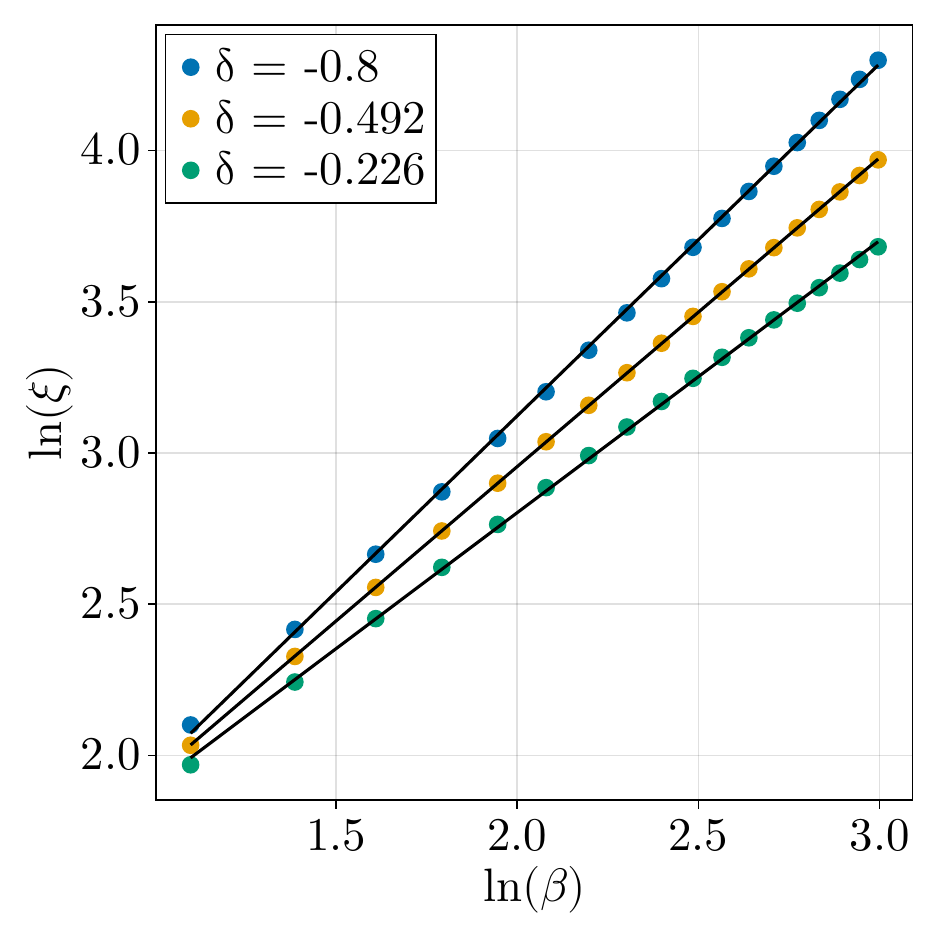} 
    \includegraphics[scale=0.4]{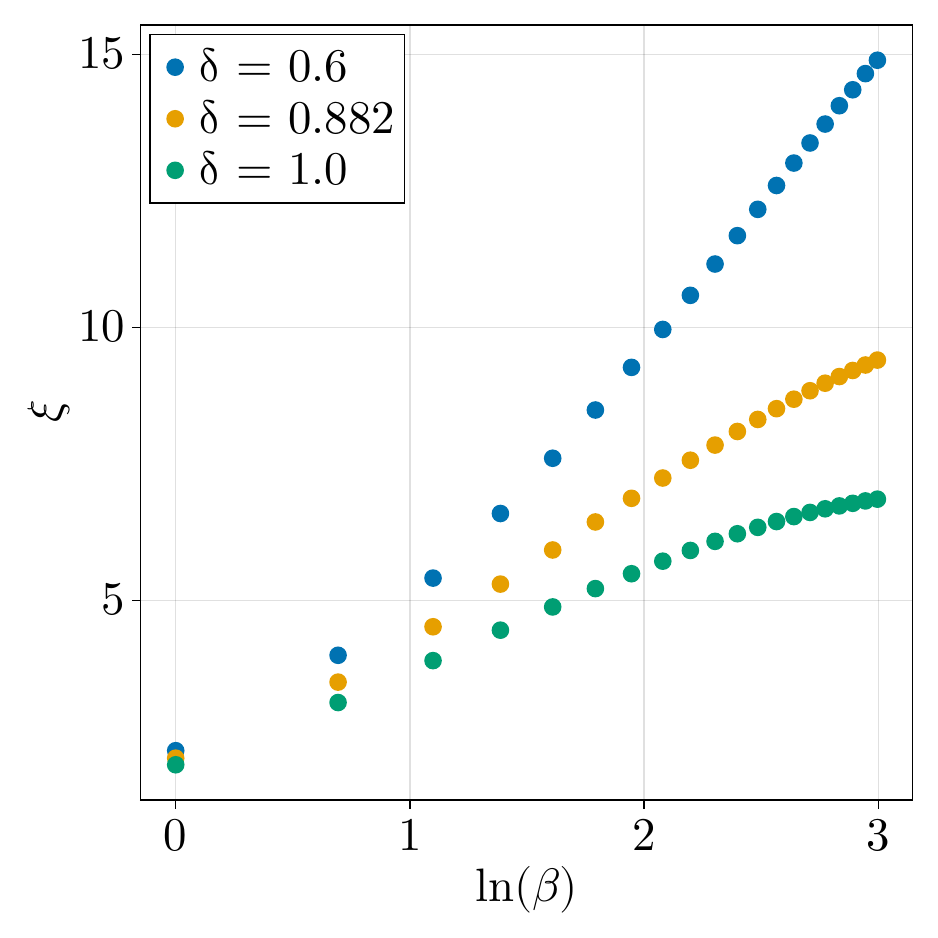} 
    \end{center}
    \caption{Left: Correlation length in function of inverse temperature for different $\delta$ on the ordered side of the quantum phase transition. The black lines represent linear fits to the data. Right: Correlation length in function of inverse temperature for different $\delta$ on the disordered side.} \label{fig:ord_dis_corr}
\end{figure}

Fig. \ref{fig:ord_dis_corr} shows in more detail the growth of the correlation length with $\ln \beta$ in both the ordered phase (left panel) and disordered phase (right panel). The results confirm the expected scaling behavior of the correlation length. In the left panel of Fig. \ref{fig:ord_dis_corr} we see that the correlation length on the ordered side grows as a power of $\beta$. Furthermore, the slope of the curves on the log-log plot changes with $\delta$, confirming the continuously varying nature of the exponent. In the right panel of Fig. \ref{fig:ord_dis_corr} the correlation length on the disordered side of the transition is shown. The data shows the expected initial growth $\propto (\ln\beta)^2$ for small $\ln \beta$. Furthermore, the correlation length at large $\beta$ increases with decreasing $\delta$ as expected.

\end{appendix}

% TODO:
% Provide your bibliography here. You have two options:

% FIRST OPTION - write your entries here directly, following the example below, including Author(s), Title, Journal Ref. with year in parentheses at the end, followed by the DOI number.
%\begin{thebibliography}{99}
%\bibitem{1931_Bethe_ZP_71} H. A. Bethe, {\it Zur Theorie der Metalle. i. Eigenwerte und Eigenfunktionen der linearen Atomkette}, Zeit. f{\"u}r Phys. {\bf 71}, 205 (1931), \doi{10.1007\%2FBF01341708}.
%\bibitem{arXiv:1108.2700} P. Ginsparg, {\it It was twenty years ago today... }, \url{http://arxiv.org/abs/1108.2700}.
%\end{thebibliography}

% \bibliographystyle{SciPost_bibstyle} % Include this style file here only if you are not using our template
\bibliography{bib.bib}

@article{igloi2005strong,
  title={Strong disorder RG approach of random systems},
  author={Igl{\'o}i, Ferenc and Monthus, C{\'e}cile},
  journal={Phys. Rep.},
  volume={412},
  number={5-6},
  pages={277--431},
  year={2005},
  url={https://www.sciencedirect.com/science/article/abs/pii/S0370157305001092?via%3Dihub}
}

@article{petHo2025random,
  title={Random transverse and longitudinal field Ising chains},
  author={Pet{\H{o}}, Tam{\'a}s and Igl{\'o}i, Ferenc and Kov{\'a}cs, Istv{\'a}n A},
  journal={J. Stat. Mech. Theory Exp.},
  volume={2025},
  number={9},
  pages={094001},
  year={2025},
  url={https://iopscience.iop.org/article/10.1088/1742-5468/adfe5a}
}

@article{hubig-time-dependent-2019,
	title={{Time-dependent study of disordered models with infinite projected entangled pair states}},
	author={Hubig, Claudius and Cirac, J. Ignacio},
	journal={SciPost Phys.},
	volume={6},
	pages={031},
	year={2019},
	publisher={SciPost},
	doi={10.21468/SciPostPhys.6.3.031},
	url={https://scipost.org/10.21468/SciPostPhys.6.3.031},
}

@article{White1992,
  title = {Density matrix formulation for quantum renormalization groups},
  author = {White, Steven R.},
  journal = {Phys. Rev. Lett.},
  volume = {69},
  issue = {19},
  pages = {2863--2866},
  numpages = {0},
  year = {1992},
  month = {Nov},
  publisher = {American Physical Society},
  doi = {10.1103/PhysRevLett.69.2863},
  url = {https://link.aps.org/doi/10.1103/PhysRevLett.69.2863}
}

@article{Vidal2003,
  title = {Efficient Classical Simulation of Slightly Entangled Quantum Computations},
  author = {Vidal, Guifr\'e},
  journal = {Phys. Rev. Lett.},
  volume = {91},
  issue = {14},
  pages = {147902},
  numpages = {4},
  year = {2003},
  month = {Oct},
  publisher = {American Physical Society},
  doi = {10.1103/PhysRevLett.91.147902},
  url = {https://link.aps.org/doi/10.1103/PhysRevLett.91.147902}
}

@Article{Vanderstraeten2019,
	title={{Tangent-space methods for uniform matrix product states}},
	author={Laurens Vanderstraeten and Jutho Haegeman and Frank Verstraete},
	journal={SciPost Phys. Lect. Notes},
	pages={7},
	year={2019},
	publisher={SciPost},
	doi={10.21468/SciPostPhysLectNotes.7},
	url={https://scipost.org/10.21468/SciPostPhysLectNotes.7},
}

@Article{Vanhecke2021,
	title={{Tangent-space methods for truncating uniform MPS}},
	author={Bram Vanhecke and Maarten Van Damme and Jutho Haegeman and Laurens Vanderstraeten and Frank Verstraete},
	journal={SciPost Phys. Core},
	volume={4},
	pages={004},
	year={2021},
	publisher={SciPost},
	doi={10.21468/SciPostPhysCore.4.1.004},
	url={https://scipost.org/10.21468/SciPostPhysCore.4.1.004},
}

@article{WuMcCoy,
  title = {Theory of a Two-Dimensional Ising Model with Random Impurities. I. Thermodynamics},
  author = {McCoy, Barry M. and Wu, Tai Tsun},
  journal = {Phys. Rev.},
  volume = {176},
  issue = {2},
  pages = {631--643},
  numpages = {0},
  year = {1968},
  month = {Dec},
  publisher = {American Physical Society},
  doi = {10.1103/PhysRev.176.631},
  url = {https://link.aps.org/doi/10.1103/PhysRev.176.631}
}

@article{Fisher1992,
  title = {Random transverse field Ising spin chains},
  author = {Fisher, Daniel S.},
  journal = {Phys. Rev. Lett.},
  volume = {69},
  issue = {3},
  pages = {534--537},
  numpages = {0},
  year = {1992},
  month = {Jul},
  publisher = {American Physical Society},
  doi = {10.1103/PhysRevLett.69.534},
  url = {https://link.aps.org/doi/10.1103/PhysRevLett.69.534}
}

@article{Fisher1995,
  title = {Critical behavior of random transverse-field Ising spin chains},
  author = {Fisher, Daniel S.},
  journal = {Phys. Rev. B},
  volume = {51},
  issue = {10},
  pages = {6411--6461},
  numpages = {0},
  year = {1995},
  month = {Mar},
  publisher = {American Physical Society},
  doi = {10.1103/PhysRevB.51.6411},
  url = {https://link.aps.org/doi/10.1103/PhysRevB.51.6411}
}

@article{Laflorencie2004,
  title = {Crossover effects in the random-exchange spin-$\frac{1}{2}$ antiferromagnetic chain},
  author = {Laflorencie, Nicolas and Rieger, Heiko and Sandvik, Anders W. and Henelius, Patrik},
  journal = {Phys. Rev. B},
  volume = {70},
  issue = {5},
  pages = {054430},
  numpages = {11},
  year = {2004},
  month = {Aug},
  publisher = {American Physical Society},
  doi = {10.1103/PhysRevB.70.054430},
  url = {https://link.aps.org/doi/10.1103/PhysRevB.70.054430}
}

@article{Schollwock2005,
  title = {The density-matrix renormalization group},
  author = {Schollw\"ock, U.},
  journal = {Rev. Mod. Phys.},
  volume = {77},
  issue = {1},
  pages = {259--315},
  numpages = {0},
  year = {2005},
  month = {Apr},
  publisher = {American Physical Society},
  doi = {10.1103/RevModPhys.77.259},
  url = {https://link.aps.org/doi/10.1103/RevModPhys.77.259}
}

@article{White2018,
  title = {Quantum dynamics of thermalizing systems},
  author = {White, Christopher David and Zaletel, Michael and Mong, Roger S. K. and Refael, Gil},
  journal = {Phys. Rev. B},
  volume = {97},
  issue = {3},
  pages = {035127},
  numpages = {14},
  year = {2018},
  month = {Jan},
  publisher = {American Physical Society},
  doi = {10.1103/PhysRevB.97.035127},
  url = {https://link.aps.org/doi/10.1103/PhysRevB.97.035127}
}

@article{VanderstraetenCluster_2021,
  title = {Symmetric cluster expansions with tensor networks},
  author = {Vanhecke, Bram and Vanderstraeten, Laurens and Verstraete, Frank},
  journal = {Phys. Rev. A},
  volume = {103},
  issue = {2},
  pages = {L020402},
  numpages = {5},
  year = {2021},
  month = {Feb},
  publisher = {American Physical Society},
  doi = {10.1103/PhysRevA.103.L020402},
  url = {https://link.aps.org/doi/10.1103/PhysRevA.103.L020402}
}

@ARTICLE{VanheckeCluster_2021,
       author = {{Vanhecke}, Bram and {Devoogdt}, David and {Verstraete}, Frank and {Vanderstraeten}, Laurens},
        title = "{Simulating thermal density operators with cluster expansions and tensor networks}",
      journal = {arXiv e-prints},
         year = 2021,
        month = dec,
          eid = {arXiv:2112.01507},
          doi = {10.48550/arXiv.2112.01507},
       eprint = {2112.01507},
}

@article{Hastings2006,
  title = {Solving gapped Hamiltonians locally},
  author = {Hastings, M. B.},
  journal = {Phys. Rev. B},
  volume = {73},
  issue = {8},
  pages = {085115},
  numpages = {13},
  year = {2006},
  month = {Feb},
  publisher = {American Physical Society},
  doi = {10.1103/PhysRevB.73.085115},
  url = {https://link.aps.org/doi/10.1103/PhysRevB.73.085115}
}

@article{Molnar2015,
  title = {Approximating Gibbs states of local Hamiltonians efficiently with projected entangled pair states},
  author = {Molnar, Andras and Schuch, Norbert and Verstraete, Frank and Cirac, J. Ignacio},
  journal = {Phys. Rev. B},
  volume = {91},
  issue = {4},
  pages = {045138},
  numpages = {11},
  year = {2015},
  month = {Jan},
  publisher = {American Physical Society},
  doi = {10.1103/PhysRevB.91.045138},
  url = {https://link.aps.org/doi/10.1103/PhysRevB.91.045138}
}

@article{Verstraete2008,
author = {F. Verstraete and V. Murg and J.I. Cirac},
title = {Matrix product states, projected entangled pair states, and variational renormalization group methods for quantum spin systems},
journal = {Advances in Physics},
volume = {57},
number = {2},
pages = {143--224},
year = {2008},
publisher = {Taylor \& Francis},
doi = {10.1080/14789940801912366},
}

@article{Roberts2017,
  title = {Implementation of rigorous renormalization group method for ground space and low-energy states of local Hamiltonians},
  author = {Roberts, Brenden and Vidick, Thomas and Motrunich, Olexei I.},
  journal = {Phys. Rev. B},
  volume = {96},
  issue = {21},
  pages = {214203},
  numpages = {12},
  year = {2017},
  month = {Dec},
  publisher = {American Physical Society},
  doi = {10.1103/PhysRevB.96.214203},
  url = {https://link.aps.org/doi/10.1103/PhysRevB.96.214203}
}

@article{Roberts2021,
  title = {Infinite randomness with continuously varying critical exponents in the random XYZ spin chain},
  author = {Roberts, Brenden and Motrunich, Olexei I.},
  journal = {Phys. Rev. B},
  volume = {104},
  issue = {21},
  pages = {214208},
  numpages = {26},
  year = {2021},
  month = {Dec},
  publisher = {American Physical Society},
  doi = {10.1103/PhysRevB.104.214208},
  url = {https://link.aps.org/doi/10.1103/PhysRevB.104.214208}
}

@article{Paredes2005,
  title = {Exploiting Quantum Parallelism to Simulate Quantum Random Many-Body Systems},
  author = {Paredes, B. and Verstraete, F. and Cirac, J. I.},
  journal = {Phys. Rev. Lett.},
  volume = {95},
  issue = {14},
  pages = {140501},
  numpages = {4},
  year = {2005},
  month = {Sep},
  publisher = {American Physical Society},
  doi = {10.1103/PhysRevLett.95.140501},
  url = {https://link.aps.org/doi/10.1103/PhysRevLett.95.140501}
}

@article{Banuls2023,
   author = "Banuls, Mari Carmen",
   title = "Tensor Network Algorithms: A Route Map", 
   journal= "Annual Review of Condensed Matter Physics",
   year = "2023",
   volume = "14",
   number = "Volume 14, 2023",
   pages = "173-191",
   doi = "https://doi.org/10.1146/annurev-conmatphys-040721-022705",
   url = "https://www.annualreviews.org/content/journals/10.1146/annurev-conmatphys-040721-022705",
   publisher = "Annual Reviews",
   issn = "1947-5462",
   type = "Journal Article",
  }

@article{Haegeman2017,
   author = "Haegeman, Jutho and Verstraete, Frank",
   title = "Diagonalizing Transfer Matrices and Matrix Product Operators: A Medley of Exact and Computational Methods", 
   journal= "Annual Review of Condensed Matter Physics",
   year = "2017",
   volume = "8",
   number = "Volume 8, 2017",
   pages = "355-406",
   doi = "https://doi.org/10.1146/annurev-conmatphys-031016-025507",
   url = "https://www.annualreviews.org/content/journals/10.1146/annurev-conmatphys-031016-025507",
   publisher = "Annual Reviews",
   issn = "1947-5462",
   type = "Journal Article",
  }

@article{Ma1979,
  title = {Random Antiferromagnetic Chain},
  author = {Ma, Shang-keng and Dasgupta, Chandan and Hu, Chin-kun},
  journal = {Phys. Rev. Lett.},
  volume = {43},
  issue = {19},
  pages = {1434--1437},
  numpages = {0},
  year = {1979},
  month = {Nov},
  publisher = {American Physical Society},
  doi = {10.1103/PhysRevLett.43.1434},
  url = {https://link.aps.org/doi/10.1103/PhysRevLett.43.1434}
}

@article{Dasgupta1980,
  title = {Low-temperature properties of the random Heisenberg antiferromagnetic chain},
  author = {Dasgupta, Chandan and Ma, Shang-keng},
  journal = {Phys. Rev. B},
  volume = {22},
  issue = {3},
  pages = {1305--1319},
  numpages = {0},
  year = {1980},
  month = {Aug},
  publisher = {American Physical Society},
  doi = {10.1103/PhysRevB.22.1305},
  url = {https://link.aps.org/doi/10.1103/PhysRevB.22.1305}
}

@article{oseledets1968multiplicative,
  title={A multiplicative ergodic theorem. Characteristic Ljapunov, exponents of dynamical systems},
  author={Oseledets, Valery Iustinovich},
  journal={Trudy Moskovskogo Matematicheskogo Obshchestva},
  volume={19},
  pages={179--210},
  year={1968},
  publisher={Moscow Mathematical Society}
}

@article{Bhatt1982,
  title = {Scaling Studies of Highly Disordered Spin-\textonehalf{} Antiferromagnetic Systems},
  author = {Bhatt, R. N. and Lee, P. A.},
  journal = {Phys. Rev. Lett.},
  volume = {48},
  issue = {5},
  pages = {344--347},
  numpages = {0},
  year = {1982},
  month = {Feb},
  publisher = {American Physical Society},
  doi = {10.1103/PhysRevLett.48.344},
  url = {https://link.aps.org/doi/10.1103/PhysRevLett.48.344}
}

@article{young1997finite,
  title={Finite-temperature and dynamical properties of the random transverse-field Ising spin chain},
  author={Young, AP},
  journal={Physical Review B},
  volume={56},
  number={18},
  pages={11691},
  year={1997},
  publisher={APS}
}

@article{young1996numerical,
  title={Numerical study of the random transverse-field Ising spin chain},
  author={Young, A Peter and Rieger, Heiko},
  journal={Physical Review B},
  volume={53},
  number={13},
  pages={8486},
  year={1996},
  publisher={APS}
}

@article{haegeman-post-2013,
  title = {Post-matrix product state methods: To tangent space and beyond},
  author = {Haegeman, Jutho and Osborne, Tobias J. and Verstraete, Frank},
  journal = {Phys. Rev. B},
  volume = {88},
  issue = {7},
  pages = {075133},
  numpages = {35},
  year = {2013},
  month = {Aug},
  publisher = {American Physical Society},
  doi = {10.1103/PhysRevB.88.075133},
  url = {https://link.aps.org/doi/10.1103/PhysRevB.88.075133}
}

@article{zauner-stauber-variational-2018,
  title = {Variational optimization algorithms for uniform matrix product states},
  author = {Zauner-Stauber, V. and Vanderstraeten, L. and Fishman, M. T. and Verstraete, F. and Haegeman, J.},
  journal = {Phys. Rev. B},
  volume = {97},
  issue = {4},
  pages = {045145},
  numpages = {31},
  year = {2018},
  month = {Jan},
  publisher = {American Physical Society},
  doi = {10.1103/PhysRevB.97.045145},
  url = {https://link.aps.org/doi/10.1103/PhysRevB.97.045145}
}

@book{xiang-density-book-2023,
  title={Density Matrix and Tensor Network Renormalization},
  author={Xiang, Tao},
  url = {https://www.cambridge.org/be/universitypress/subjects/physics/condensed-matter-physics-nanoscience-and-mesoscopic-physics/density-matrix-and-tensor-network-renormalization?format=HB},
  year={2023},
  publisher={Cambridge University Press}
}

\nolinenumbers

\end{document}